\documentclass[11pt]{article}
\usepackage{graphicx}
\usepackage{bm}
\usepackage{subfigure}
\usepackage{natbib}
\usepackage{amssymb}
\usepackage[tbtags]{amsmath}
\usepackage{amsfonts}
\usepackage{layout}
\usepackage{setspace}
\usepackage{psfrag}
\usepackage{epstopdf}
\usepackage[colorlinks,citecolor=blue,linkcolor=blue]{hyperref}
\usepackage[all]{hypcap}
\usepackage{morefloats}
\usepackage{xcolor}
\usepackage{appendix}

\newcommand{\pe}[1]{$Pe=#1$}
\newcommand{\ph}[1]{$\phi=#1$} 

\newcommand{\ar}[1]{$AR=#1$} 
\newcommand{\nvec}{$\mathbf{n}\;$}

\setcitestyle{super,comma,open={},close={}}

\begin{document}

\title{Orientation and microstructure in sheared Brownian suspensions of anisotropic dicolloidal particles}
\author{Amit Kumar$^\dagger$ and Jonathan J. L. Higdon$^*$ \\
Department of Chemical and Biomolecular Engineering\\
University of Illinois at Urbana-Champaign, Urbana, IL 61801, USA}
\date{\today}
\maketitle

\let\thefootnote\relax\footnotetext{$^\dagger$ Present address: Department of Chemical and Biological Engineering, University of Wisconsin-Madison}
\let\thefootnote\relax\footnotetext{$^*$Corresponding author. Email: jhigdon@illinois.edu}

\begin{abstract}
Orientation and microstructure are investigated in sheared  Brownian suspensions of 
hard dicolloidal particles, with the dicolloids modeled as two fused spheres of
varying radii and center to center separations. Two different particle shapes 
named homonuclear (aspect ratio 1.1) and fused-dumbbells (aspect ratio 1.5) were
considered. Hydrodynamic interactions between the particles were computed
with a modified lubrication model called Fast Lubrication Dynamics. Studies
were conducted for a wide range of volume fractions between $0.3 \leq \phi \leq 0.5$ 
and P\`{e}clet numbers between $0 \leq Pe \leq 1000$. The microstructure was 
found to be disordered at all volume fractions, though signatures of weak string
like ordering were  evident particularly in $\phi=0.5$ homonuclear suspensions 
at intermediate to high shear rates ($Pe$ in the range $10-100$). 
Complex orientation behavior was observed as a function of shape, shear rates, and volume fractions.
At very low shear rates, random orientation distribution was observed in all 
cases. At the highest shear rates, orientation distribution in suspensions of
homonuclear particles exhibited a shift towards an alignment with the  vorticity axis at all volume fractions,
while in suspensions of fused-dumbbells it exhibited a shift away from the vorticity axis at low volume 
fractions and a negligible shift at higher volume fractions. 
The orientation behavior is further characterized by examining the orientation
distribution in the velocity--gradient plane -- in this case an increased  particle 
alignment with the velocity axis is generally observed with increasing 
volume fractions, but not universally with increasing shear rates. Mechanistic
origins for the complex orientation behavior as a function of shear rate, volume fraction, 
and particle shape is described.
\end{abstract} 

\section{Introduction}\label{sec:intro}
\begin{figure}
\centering
\includegraphics[width=0.5\textwidth]{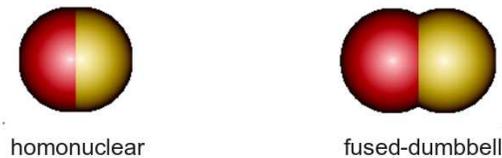}
\caption{Examples of dicolloidal particles investigated in the current work. 
	The two particle shapes are named homonuclear and fused-dumbbell particles as noted in the figure.}
\label{fig:shapes}
\end{figure}

In recent times anisotropic particles have attracted considerable interest
in a wide range of fields. This surge in interest is arguably fueled by 
the recent synthesis of a variety of novel particle shapes \citep{blaaderen06, yang08},
which themselves are enabled by the recent advances in particle synthesis techniques.
Sizes of these particles are usually in the nanometer range with O(100) nm being a typical size.
Many applications have been proposed for these anisotropic particles;
 some prominent examples include their assembly into advanced
 nanostructured materials \citep{glotzer07} and for drug delivery
\citep{mitragotri08, kumar_rev} among others.
An interesting class of recently synthesized anisotropic particles are the  dicolloidal particles
\citep{blaaderen05, mock06, weitz06}, whose geometry is closely
approximated by the union of two intersecting spheres of varying
radii and center to center separations. Figure  (\ref{fig:shapes})
illustrates several dicolloidal particle shapes examined in the present effort.
A significant feature of dicolloidal particles is its low aspect
ratio (AR) with an upper bound of 2.0. Owing to this low aspect
ratio, mildly aspherical dicolloids often exhibit microstructural
and rheological behavior similar to that in suspensions of
spheres. However, in addition to the positional microstructure,
suspensions of dicolloids possess distinct orientational
microstructure. The goals of present work are to characterize 
both the orientational and the positional microstructure
in sheared suspensions of \textit{Brownian} dicolloidal particles. 
This continues our previous work on dicolloidal particles where we investigated the
orientational and positional microstructure in sheared \textit{non-Brownian} suspensions \citep{kumar_jor}.
We review some of the important observations in this study  below.
In addition, we review below other pertinent literature on the orientation behavior 
in sheared suspensions of anisotropic particles.

\begin{figure}
\centering
\includegraphics[width=0.4\textwidth]{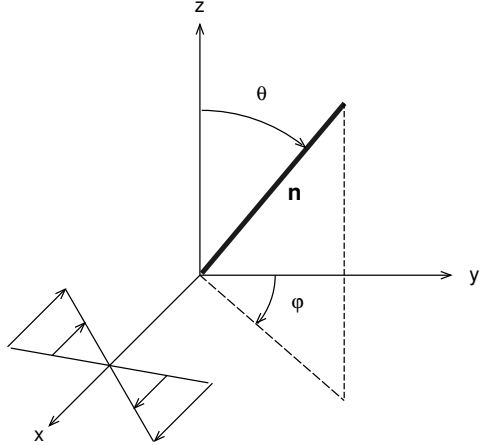}
\caption{Definition of the spherical polar coordinate $\theta$ and the azimuthal angle $\varphi$.}\label{fig:defn}
\end{figure}

In the absence of non-hydrodynamic effects (such as Brownian motion or electrostatic forces),
\citet{jeffery22} showed that a single spheroid (or more generally
an ellipsoid) in a linear shear flow would rotate indefinitely in
a single parameter family of closed orbits, commonly known as the
Jeffery orbits. The parameter characterizing an orbit is called
the orbit constant $C$ whose value varies from $C=0$ for particle director 
aligned with the vorticity axis to $C=\infty$ for particle alignment in the velocity-gradient
plane. For an axisymmetric body, such as a spheroid or a dicolloid, its orientation may be
specified by spherical coordinates ($\varphi,\theta$) where
$\theta$ measures the angle between the particle director and $z$ axis, while
$\varphi$ is angle between the director and the $y$ axis measured clockwise
in the  $x-y$ plane (see Fig. \ref{fig:defn}). The evolution of the orientational angles
($\varphi,\theta$) with time $t$ is given in terms of the orbit constant $C$ by
\begin{subequations}\label{eq:jeff}
\begin{equation}\label{eq:jeffa}
\tan\varphi = r_s \tan \left(\displaystyle \frac{\dot{\gamma} t}{r_s + 1/r_s} + \kappa \right),
\end{equation}
\begin{equation}\label{eq:jeffb}
\tan \theta =  \displaystyle \frac{C r_s }{ \sqrt{r_s^2\cos^2 \varphi + \sin^2 \varphi}},
\end{equation}
\end{subequations}
where $\dot{\gamma}$ is the shear rate, $r_s$ is the aspect ratio of the spheroid, while 
$\kappa$ denotes the initial phase in the orbit. A particularly simple limit is that of
a spherical particle ($r_s=1$), in which case Jeffery's orbits corresponds to lines of 
constant $\theta$ (more specifically, $\theta =\tan^{-1} C$). Later, \citet{bretherton62} showed that
in a linear shear flow most axisymmetric bodies, which includes dicolloids,
will also rotate in Jeffery orbits with $r_s$ in the equations above replaced by
an effective aspect ratio $r_e$ of the particle. It can be shown that the rate
of rotation of a particle in any given orbit is non-uniform, such that its director
spends a greater fraction of the time aligned in the velocity--vorticity 
plane in comparison to the gradient--vorticity plane. As such, under shear, a single 
non-Brownian anisotropic particle by itself will exhibit an enhanced time-averaged flow 
alignment (parallel to the velocity axis) in conjunction with a reduced gradient alignment (henceforth, 
by particle alignment we imply particle director alignment). On the other hand, the
time-averaged vorticity alignment is mostly determined by the initial 
orientation of the particle; the initial orientation sets the orbit constant $C$, 
and the vorticity alignment does not change appreciably with time in any given orbit  for 
particles with small degree of anisotropy.

The effect of Brownian motion on the orientation behavior of  
axisymmetric particles have been characterized in
several theoretical studies in the limit of infinite dilution, i.e., non-interacting particle suspensions. 
\citet{leal71} investigated the orientation distribution
in suspensions of spheroids in the limit of very weak Brownian motion.
The key argument in this study was that the rotation of a particle in any given
Jeffery's orbit is unaltered when the Brownian motion is very weak. However,
after sufficiently long times, the first effect of the weak Brownian
motion will be to yield steady state distributions across orbits
independent of the initial state of the suspension.
Since the rate of rotation of a particle in any given orbit is unchanged, 
this result implies an enhanced particle alignment with the velocity axis and
a reduced particle alignment with the gradient axis, say, in comparison to an
isotropic distribution. In an earlier study \citep{kumar_jor}, we showed that
in the limit of weak Brownian motion, the  particle alignment with the vorticity
axis remains approximately isotropic for dicolloids.  \citet{leal72} 
later extended their previous analysis to include higher order approximations, though
still in the weak Brownian limit. This study, in comparison to Jeffery's 
solution, showed an enhanced probability at orientations just prior to complete flow alignment, 
in conjunction with a reduced probability at orientations just after complete flow alignment.
This behavior results from the fact that the Brownian motion causes diffusion away from the highly
populated flow aligned states, which, to first approximation, occurs with equal probability towards
the pre flow aligned state and post flow aligned state. The advection, on the other hand, causes particle
motion in a single direction of the undisturbed Jeffery's orbit. The balance between these two effects
leads to the aforementioned probability distribution.

In the limit of strong Brownian motion, the orientation distribution in non-interacting
suspensions of axisymmetric particles is provided by, e.g., \citet{brenner74}.
At equilibrium, as could be expected, the orientation distribution was found to be 
isotropic.  The first effect of the weak flow is to enhance the orientation
probability around the extensional axis of the 
shear flow and a reduced probability along its compressional axis.
The effect of shear flow is understood by noting that the overall 
flow in this case can be expressed as a sum of an extensional component and a rotational component.
The rotational component has no effect at the leading order as the rate of rotation
of the particle due to this component is uniform and the initial distribution, i.e., at equilibrium, isotropic.
The effect of the extensional component of the flow is to rotate the particle away from the 
compressional axis towards alignment with the extensional axis, which explains
the above observation. At higher order approximations, both the extensional as
well as the rotational component of the flow will contribute. The effect of the
rotational component of the flow will be to rotate the orientation distribution 
clockwise, thereby bringing the probability maxima closer to the velocity 
axis and the probability minima closer to the gradient axis. At 
very high shear rates, it can be expected that the weak Brownian limit discussed
above will be reached, i.e., the region of maximum orientation probability
will coincide with the velocity axis, while the region of minimum 
orientation probability will coincide with the gradient axis.

There have been relatively few studies on the orientation behavior in Brownian 
suspensions of anisotropic particles at finite concentrations. Some are 
discussed next. \citet{higdon08b} investigated the orientation behavior
in concentrated Brownian suspensions of plate-like 
particles ($AR$ between 3 and 7) via an extension of the Stokesian Dynamics
technique. Their studies indicate an increasing particle alignment
in the velocity-vorticity plane with increasing shear rates, volume fractions
and aspect ratios.
\citet{wagner06} experimentally measured flow alignment in suspensions of
ellipsoids of comparatively lower aspect ratios ($AR$ between 2 and 7).
This study showed that the flow alignment (parallel to velocity
axis) was preferred at all shear rates with the maximum in flow
alignment coinciding with the onset of shear thickening.
At even higher shear rates,  a decrease in flow
alignment was observed. This study also indicated an increased
flow alignment with higher particle loadings as well as aspect ratios.
The authors attributed the decrease in flow alignment at high shear rates to the
formation of hydroclusters, which will have a smaller degree of anisotropy than the 
constituent particle itself; this reduction in the effective anisotropy explains the 
decrease in particle flow alignment.
Besides the studies noted above, there have also been many other studies
on the orientation behavior in suspensions of high aspect ratio
non-Brownian fibers, which is of limited interest here. For a detailed survey of 
such suspensions  the reader is referred to our earlier publication  \citep{kumar_jor}.

We recently investigated the orientation behavior in sheared non-Brownian suspensions
of dicolloidal particles using the Stokesian dynamics technique \citep{kumar_jor}.
This study explored the effect of particle shape as well as volume fraction. Two of the particle shapes investigated
in this study are shown in Fig. (\ref{fig:shapes}).
As noted in the figure, the two particles shapes, both of which are symmetric dicolloids,
are named homonuclear and fused-dumbbell particles. The geometric aspect ratio  of 
these two particles are \ar{1.1} and \ar{1.5}, respectively.
In this this study it was shown that, at low volume fractions, both the particle shapes
exhibited an orientation drift away from the vorticity axis towards
the velocity--gradient plane. This drift was considerably higher for the fused-dumbbell
particle than the homonuclear particle. The origin of this behavior was attributed
to non-uniform fluctuations in the orientation space -- particles aligned with 
the vorticity axis were found to undergo stronger fluctuations in its orientation
as it underwent collisions with other particles. 
As discussed in \citet{koch88}, non-uniform fluctuations in the orientation space will result in a
drift away from orientations with stronger fluctuations in agreement
with the observations. At higher volume fractions, an orientation drift towards 
the vorticity axis away from the velocity--gradient plane was observed. 
The drift towards the vorticity axis was higher for the homonuclear
particles than for the fused-dumbbell particles. In suspensions of force free and torque 
free particles, as is approximately the case in non-Brownian  suspensions with a very short
range repulsive force, the particle angular velocity can be directly related to the hydrodynamic stresslets,
which refers to the symmetric part of the first moment of the fluid force on the particle. 
Indeed, it was shown that the particle drift towards the vorticity
axis was correlated with the hydrodynamic stresslet, in particular with the 
second normal difference component of the stresslet \citep{kumar_jor}. In non-Brownian suspensions, the second
normal stress differences are negative \citep{kumar_jor,brady2002}, which 
was shown to drive the orientation distribution towards an enhanced vorticity alignment.
It is interesting to note that, in a non-Newtonian second order fluid, 
\citet{leal75} showed that a high aspect ratio fiber subjected to shear will also
drift towards vorticity alignment if second normal stress coefficient of the fluid is negative.

In the present effort, we will conduct numerical simulations for
the dynamics of sheared suspensions of Brownian dicolloids.
The hydrodynamic interactions as well as the Brownian motion
of the particle are computed using the Fast Lubrication
Dynamics (FLD) algorithm \citep{kumar10_pre,kumar10_dis}. The FLD
algorithm is an approximation of the well known Stokesian Dynamics technique \citep{brady2001,kumar_jfm}.
In the FLD algorithm, the overall resistance tensor is expressed as the sum of a pairwise lubrication term and 
a novel isotropic resistance term chosen so as to match the mean
particle mobility from the more detailed Stokesian Dynamics.  
Using this algorithm we investigate the orientation and microstructure
in Brownian hard dicolloidal particle suspensions as a function of shear rate and volume fraction.
Two different particle shapes are considered here: the homonuclear and the 
fused-dumbbell particle. The microstructure is found to lack any long
range order in the range of volume fraction investigated ($0.3 \leq \phi \leq 0.5$),
though hints of weak string like ordering becomes evident at the highest volume 
fraction over a range of shear rates, especially in suspensions of homonuclear particles.
The particle orientation is found to be random at low shear rates in all cases. 
At the highest shear rates, orientation distribution in suspensions of
homonuclear particles exhibits a shift towards an alignment with the  vorticity
axis at all volume fractions, while in suspensions of fused-dumbbells it exhibits
a shift away from the vorticity axis at low volume. 
The orientation distribution in the velocity--gradient plane shows an 
increased flow alignment with increasing volume fractions for both the
particle shapes. The flow alignment is also found to increase with increasing
shear rates at lower volume fractions, though, at higher volume fractions,
this is not necessarily the case. The mechanism responsible for these orientation
distributions with shear rate, volume fraction, and particle shape is described
in light of previous studies on non-Brownian and strongly Brownian suspensions.

The organization of this article is as follows. In Sec. (\ref{sec:form}) we present 
the Fast Lubrication Dynamics technique for computing the hydrodynamic and Brownian
forces on the particles. The solution procedure for computing
	the unknown  velocities and angular velocities of the particles is also discussed here.
	Next, in Sec. (\ref{sec:struct}), the results for the microstructure as a function
	of volume fraction, shear rate and particle shape is presented.
Following this,  the orientation behavior is characterized in Sec. (\ref{sec:oc}).
Concluding remarks are lastly presented in Sec. (\ref{sec:conc}).

\section{Formulation}\label{sec:form}

The motion of Brownian particles in a viscous fluid can be described by Newton's
second law written as
\begin{equation} \label{eq:newton}
\mathbf{m} \cdot
\frac{d{\mathbf{U}}}{d{t}}=\mathbf{F}^H+ \mathbf{F}^{B} + \mathbf{F}^P,
\end{equation}
where $\mathbf{m}$ is the mass/moment of inertia tensor of the
particles, $\mathbf{U}$ is the generalized velocity/angular
velocity vector, while $\mathbf{F}^H$, $\mathbf{F}^B$, and $\mathbf{F}^P$ are the
generalized force/torque vectors arising respectively from the hydrodynamic
stress in the fluid, stochastic Brownian motion, and interparticle interactions.
Here the vectors $\mathbf{U}$, $\mathbf{F}^H$, $\mathbf{F}^B$ and
$\mathbf{F}^P$ are $6N_p$ vectors where $N_p$ is the number of
particles in the system; the mass/moment of inertia
$\mathbf{m}$ is a $6N_p$ square block diagonal matrix where each
$6 \times 6$ block consists of the mass and moment of inertia
tensor of each individual particle. We assume that the particle
size is small, and the inertial effects are negligible. With this assumption, the left
hand side of \eqref{eq:newton} is zero, and the sum of the
forces and torques on each individual particle must be zero at
every instant of time:
\begin{equation} \label{eq:newton1}
\mathbf{F}^H+ \mathbf{F}^{B} + \mathbf{F}^P = 0.
\end{equation}
We discuss next the numerical calculation of the above three type of forces, i.e.,
$\mathbf{F}^{H}$, $\mathbf{F}^{B}$, and $\mathbf{F}^{P}$.

\subsection{Hydrodynamic force ($\mathbf{F}^H$)}
Owing to its small size, the fluid motion around the particles is
governed by the Stokes equations. In these low Reynolds number flows, the hydrodynamic
force on each particle is a linear function of the fluid
velocity and may be determined from the solution of the
governing equations for a specified set of particle
configurations, particle velocities and of a prescribed
undisturbed flow field. The linear relationship may be expressed
in terms of a configuration dependent $N_p$ body resistance tensor
$\mathbf{R}$ as follows:
\begin{equation}\label{eq:FH}
\left( \begin{array}{c} \mathbf{F}^H \\ \mathbf{S}^H \end{array}
\right) = \mathbf{R} \cdot \left( \begin{array}{c}
\mathbf{U}^{\infty}-\mathbf{U} \\ \mathbf{E}^{\infty}
\end{array} \right).
\end{equation}
In the above equation, a homogeneous undisturbed linear shear flow is
assumed which determines the generalized velocity/angular
velocity $\mathbf{U}^{\infty}$ evaluated at the center of each
particle and the uniform rate of strain tensor denoted by
$\mathbf{E}^{\infty}$. $\mathbf{F}^H$ is the usual force/torque
$6N_p$ vector and $\mathbf{S}^H$ denotes the stresslets on the
particle, the symmetric part of the first moment of the force on
each particle, and can be expressed by a $5N_p$ vector.

In a previous effort \citep{kumar_jfm}, we developed a
Stokesian dynamics technique to approximate the resistance 
tensor $\mathbf{R}$ in suspensions of non-spherical particles including dicolloids.
In the Stokesian Dynamics approach, the total resistance tensor $\mathbf{R}$
in Eq. (\ref{eq:FH}) is expressed by the following approximation:
\begin{equation}\label{eq:RRLB}
\mathbf{R} = \mathbf{R}^{MB} + \mathbf{R}^{LB},
\end{equation}
where $\mathbf{R}^{MB}$ is a many-body resistance tensor
accurate for widely separated particles. When particles come
near contact however, the small particle gaps lead to strong
interactions with $\mathbf{R}$ diverging at contact.  The many
body resistance tensor cannot capture this singular behavior due
to the truncated multipole expansion employed in its
calculation, hence, a correction term is added to account for
the missing terms. This correction tensor denoted by
$\mathbf{R}^{LB}$  is based on asymptotic lubrication theory for
nearly touching particles. Unlike $\mathbf{R}^{MB}$,
$\mathbf{R}^{LB}$ is a sparse matrix as the lubrication
interactions affect only near neighbors, and most importantly,
the asymptotic lubrication contributions are pairwise additive
as the interaction is highly localized around the point of
contact. In the Stokesian Dynamics technique the bulk of the computational 
cost is associated with the calculation of the far-field many-body interactions.
However, particularly in concentrated suspensions, it is the lubrication interactions
that dominate the overall particle-particle interactions.
These considerations lead us to the development 
of the Fast Lubrication Dynamics technique (FLD), which is 
described below in the case of suspensions of spheres; its extension to anisotropic
dicolloidal particles is presented later in Sec. (\ref{sec:dicoll}).

In the FLD approach,  the overall resistance tensor is approximated as \citep{kumar10_pre,kumar10_dis}
\begin{equation}\label{eq:fld}
\mathbf{R} = \mathbf{R}_0 + \mathbf{R}_{\delta},
\end{equation}
where $\mathbf{R}_{\delta}$ is based on the near-field lubrication
interactions summed pairwise \citep{kumar10_dis,melrose97}, while a novel isotropic term
$\mathbf{R}_0$ (diagonal tensor, function of volume fraction alone) is selected to match the mean particle mobility
(equivalently the short time self-diffusivity) from the more detailed Stokesian Dynamics
technique for spheres \citep{kumar10_pre,kumar10_dis}. Somewhat related approaches have been described
before in the literature by \citet{melrose97} and by \citet{banchio03}.  The FLD approach has been
shown to produce results comparable to the Stokesian Dynamics technique, while requiring only a fraction of the
latter's cost; see \citet{kumar10_dis} and \citet{kumar10_pre} for details;
further validation is provided in the recent work of \citet{schunk12}.
We discuss next the extension of the FLD technique for modeling hydrodynamic interactions
in dicolloidal particle suspensions.

\begin{figure}
\centering
\includegraphics[width=0.5\textwidth]{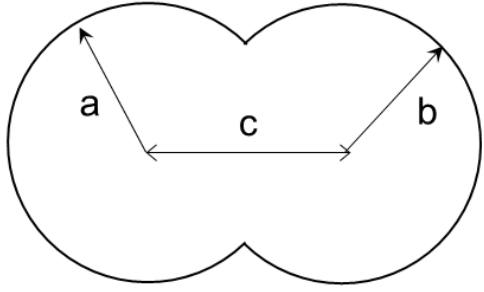}
\caption{A cartoon of a dicolloidal particle. Radii of the two 
	component spheres forming the dicolloid are $a$ and $b$, while the distance between
	the centers of the two component spheres is $c$. In the present work only
	symmetric dicolloids with $a=b$ are considered.}\label{fig:dicoll}
\end{figure}

\begin{figure}
\centering
\includegraphics[width=0.45\textwidth]{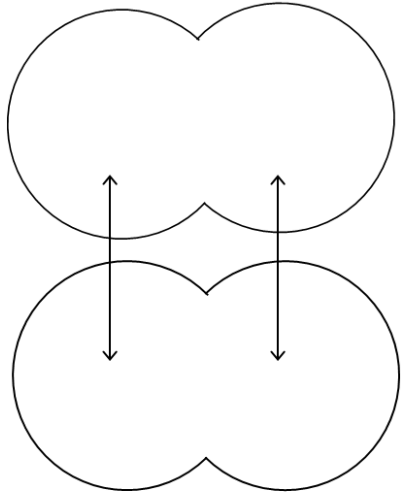}
\caption{A cartoon showing two pairwise interactions between spherical
	components forming the two dicolloids.}\label{fig:pair}
\end{figure}

\subsubsection{FLD technique for dicolloidal Particles}\label{sec:dicoll}
In the FLD technique for dicolloidal particles (Fig. \ref{fig:dicoll} shows a generic
dicolloidal particle), the overall resistance tensor $\mathbf{R}$ is expressed by an equation similar
to Eq. (\ref{eq:fld}). For simplicity, in the present effort, we take the value of 
$\mathbf{R}_0$ in a suspension of dicolloids to be the same as in a suspension
of spheres at the same volume fraction, irrespective of the exact shape of the particle
under consideration. For particles with small degree of anisotropy this is a fair assumption,
though some error in the value of $\mathbf{R}_0$ can be expected at larger degree
of anisotropy. Note that the largest aspect ratio particle investigated in the current work is
$AR=1.5$. Additionally, since our interest is only in concentrated suspensions with volume fraction $\phi \geq 0.3$, 
we estimate the contribution from this error to be small, as the hydrodynamic interactions will
be dominated by the strong lubrication forces at these high volume fractions. 

There are some additional considerations with the present choice of $\mathbf{R}_0$,
which deserve a brief comment. The tensor $\mathbf{R}_0$ in some ways represents the effective one-body resistance 
in the suspension. Our choice of $\mathbf{R}_0$ as an isotropic diagonal tensor implies that 
a dicolloidal particle, in the absence of interactions with other particles, will behave just 
like a sphere. For example, if we study an isolated dicolloidal particle with the 
current implementation of the FLD algorithm, then the dicolloid will not exhibit preferred flow alignment as might 
be expected for an anisotropic particle \citep{jeffery22,kumar_jor}. However,
this is not a great concern here as the particle dynamics, including its orientation behavior,
will be dominated by the lubrication interactions at the volume fractions considered here.

Next, we discuss the computation of the near field lubrication  term  $\mathbf{R}_{\delta}$
in suspensions of dicolloidal particles. As shown in Fig. (\ref{fig:dicoll}), a dicolloidal particle
consists of two fused spheres (only symmetric dicolloids with $a=b$ are considered here).
Given this fact, the lubrication interaction between two
dicolloids can be modeled as a pairwise sum of one or more interactions between
the constituent spheres forming the two dicolloids; this is schematically shown in  Fig. (\ref{fig:pair}) and 
discussed in detail elsewhere \citep{kumar_jfm,kumar10_dis}. Further note that 
up to four spherical node pairs can be formed between two dicolloidal particles, .
In our algorithm, we do not necessarily consider all the four pairs. The selection
of a spherical node pair for the lubrication calculation depends on two factors. 
First involves the satisfaction of a cut-off separation requirement -- only those
spherical nodes pairs are considered in the lubrication calculation which have a center
to center separation less than $2.5a$, where $a$ is the radius of the spherical node. 
Second factor involves the concept of inter-visibility -- only those spherical node
pairs are considered for the lubrication calculation which are inter-visible. 
Mathematically, if the line joining the centers of the two spherical
nodes passes through each of their surfaces, then that pair has inter-visibility and is considered for
the lubrication calculation; otherwise it is rejected. This scheme essentially prevents double counting
which can happen otherwise. We remark that similar procedures are used extensively in the literature
for computing interparticle interactions between patchy particles -- a popular example is the 
	Kern-Frenkel model  \citep{frenkel03}.

\subsection{Brownian Force ($\mathbf{F}^B$)}
Brownian forces and torques can be obtained from the fluctuation dissipation theorem and the equipartition
of energy \citep{russel89}, which dictates
\begin{subequations}\label{eq:fluc}
\begin{equation}
\langle \mathbf{F}^B \rangle =0
\end{equation}
\begin{equation}
\langle \mathbf{F}^B\mathbf{F}^B \rangle = 2kT \; \mathbf{R}_{FU}/\Delta t
\end{equation}
\end{subequations}
where $\langle \rangle$ denotes an ensemble average, $\mathbf{R}_{FU}$ denotes the component of the
resistance tensor representing generalized force-velocity coupling,
while $\Delta t$, $k$ and $T$ denote the time step employed in the algorithm,
the Boltzmann constant, and the system temperature respectively \citep{foss2000,melrose97,kumar10_pre}.
Brownian forces/torques satisfying the above equations are easily
obtained by employing uncorrelated random numbers with zero mean
and unit variance. Briefly, we begin by writing the Brownian force/torque
vector $\mathbf{F}^B$ as 
\begin{equation}
\mathbf{F}^B=\sqrt{\frac{2kT}{\Delta t}} \left( \mathbf{A}\mbox{\boldmath{$\alpha$}} + \mathbf{B}\mbox{\boldmath{$\beta$}} \right),
\label{eq:FB}
\end{equation}
where \mbox{\boldmath{$\alpha$}} and \mbox{\boldmath{$\beta$}} are $6N_p$ vectors
whose elements are uncorrelated random numbers, say $\gamma_i$, satisfying the
aforementioned properties, namely
\begin{subequations}
\begin{equation}
\langle \gamma_i \rangle =0,
\end{equation}
\begin{equation}
\langle \gamma_i \gamma_j \rangle = \delta_{ij},
\end{equation}
\end{subequations}
where $\delta_{ij}$ is the Kronecker delta function.
With the random numbers satisfying the above properties,
we only need to satisfy the following two equations to obtain the
agreement of Eq. (\ref{eq:FB}) with Eq. (\ref{eq:fluc})
\begin{equation}
\begin{array}{c}
\mathbf{AA}^T = \mathbf{R}_0^{FU} \\
\mathbf{BB}^T = \mathbf{R}_{\delta}^{FU} \\
\end{array}
\end{equation}
The above equations require one to compute the
square root of the matrices $\mathbf{R}_0^{FU}$
and $\mathbf{R}_{\delta}^{FU}$. In the present
algorithm these are easily obtained as $\mathbf{R}_0^{FU}$
is a diagonal matrix, while contributions from $\mathbf{R}_{\delta}^{FU}$
matrix can be summed pairwise. Details can be found in \citep{melrose97,kumar10_dis}.

\subsection{Repulsive Force $\mathbf{F}^P$}\label{sec:ms_ARF}
During time stepping, due to integration errors, particle overlaps can occur.
To overcome this, we employ a short range repulsive force  $\mathbf{F}^P$ between
the spherical node pairs as follows \citep{kumar_jor}:
\begin{equation}
\mathbf{F}^P = \left \{
\begin{array}{cc}
\displaystyle C_p \left(\frac{\delta_{min}}{\delta}\right)\,(\eta - \frac{1}{2} \eta^2)^3 \, \mathbf{d}  & \textnormal{if} \;\; \delta < \delta_{min} \\ \\
0  & \textnormal{if} \;\; \delta > \delta_{min}
\end{array}
\right.,
\end{equation}
where $\delta$ is the gap between the two spherical nodes, $\eta = 1-\delta/\delta_{min}$, $\mathbf{d}$ is the unit vector pointing from one spherical node to the other, and $C_p$ is a parameter described below. In this work, the 
range of the repulsive force $\delta_{min}$ was fixed at $10^{-3}a$.
For gaps smaller than a specified numerical tolerance, $\delta_{num}$, 
both the hydrodynamic force and interparticle force are capped by evaluating at
$\delta_{num}$. This tolerance for all simulations presented here was set at $\delta_{num}=10^{-5}a$.
The interparticle force between a pair of dicolloids was computed as the sum of
pairwise forces between the spherical nodes comprising the particles. The above form of
the repulsive force gives a near hard sphere behavior with nonzero repulsive interaction
only when the gap between a spherical node pair becomes less than $10^{-3}a$.

\subsection{Solution procedure}
The goal of the overall solution procedure is to compute the unknown 
particle velocities and angular velocities for a given set of Brownian forces, 
interparticle forces, and the imposed bulk  (shear) flow.
The unknown particle velocities/angular velocities can be obtained from the solution of the 
governing equation (\ref{eq:newton1}) written as:
\begin{equation}
\mathbf{R}_{FU}\cdot(\mathbf{U}^\infty-{U}) = -(\mathbf{F}^P + \mathbf{F}^B) - \mathbf{R}_{FE}\cdot \mathbf{E}^\infty.
\label{eq:sol}
\end{equation}
The above equation is obtained from Eq. (\ref{eq:newton1}) by replacing $\mathbf{F}^H$ with
its equivalent expressed in terms of the force-velocity ($\mathbf{R}_{FU}$)
and force-rate of strain ($\mathbf{R}_{FE}$) couplings.
In the present effort, we solve Eq. (\ref{eq:sol}) iteratively with the conjugate gradient algorithm \citep{saad},
which is applicable here owing to the symmetry and positive definiteness of the resistance coupling $\mathbf{R}_{FU}$.
Once the particle velocities and angular velocities have been computed, the particle positions and 
orientations are evolved in time with a third order explicit Runge-Kutta method. The choices of the time-step
$\Delta t$ employed in the time integration step as well as the total time of integration $t$ are discussed below. 
It is relevant to note here that all simulations in the present effort were initiated
with a random configuration generated by a Monte Carlo method. A Monte Carlo simulation for dicolloidal particles
involved a random displacement to both the particle's orientation as well
as its center of mass at each step, and this was continued until an apparent steady state was attained.

\begin{figure}
\centering
\includegraphics[width=0.6\textwidth]{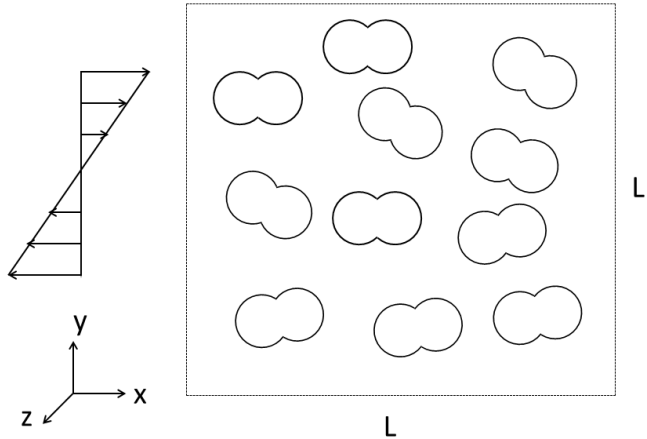}
\caption{Schematic of the system geometry: a cubic simulation box (side L) with $N_p=1000$ 
	particles subjected to a simple shear flow is considered. Periodic boundary conditions
	are employed in all three directions. }\label{fig:geom}
\end{figure}
\subsection{Simulation Parameters}\label{sec:ms_parameters}
In this section, we discuss some of important physical as well as numerical parameters 
in the current study. Perhaps the most important parameter in the present effort
is the shear rate $\dot{\gamma}$, which is expressed in terms of the non-dimensional P\`{e}clet number ($Pe$).
Note that the P\`{e}clet number gives the ratio of the diffusive and the convective time scales, 
and is defined in the present effort as: 
\begin{equation}
Pe =  \frac{\dot{\gamma} a^2}{D_0},
\end{equation}
where $a$ is the radius of the spherical node forming the dicolloid, and $D_0=kT/6\pi\mu a$ is the
Stokes-Einstein diffusivity based on the radius of the spherical node.
Alternative definitions of $Pe$ based on length scales other than $a$ is possible, though that will not change 
the value of $Pe$ appreciably given the small degree of anisotropy of dicolloids.
In the current effort, $Pe$ was varied between $0 \leq Pe \leq 1000$. 
Another important parameter in the current study is the particle volume fraction $\phi$, 
which was varied in the range $0.3 \leq \phi \leq 0.5$. Lastly, in the context of the present work,
two particle shape parameters are required for defining a dicolloid. The two shape parameters 
are the ratio of the radii of the component spheres, $b/a$, and the ratio of the
center to center distance and one of the radius, $c/a$; see Fig. (\ref{fig:dicoll})
illustrating a generic dicolloidal particle. In the present study, we have investigated
two different dicolloidal particles named as homonuclear and fused-dumbbell particles
(Fig. \ref{fig:defn}). The shape parameters (b/a,c/a) for these two particle shapes are 
(1,0.2) for homonuclear particles and (1,1) for fused-dumbbells. The geometric aspect ratio
for these particles can be obtained as $AR = (2a+c)/2a$, which yields $AR=1.1$
for homonuclear particles and $AR=1.5$ for fused-dumbbell particles.

There are several other parameters associated with the numerical solution procedure. 
First is the choice of the simulation box, which is taken to be cubic here
with periodic boundary conditions employed in all three directions. 
A schematic of the system geometry is illustrated in Fig. (\ref{fig:geom}).
The number of particles in the simulation box was set to $N_p=1000$.
The next parameter is the non-dimensional time-step $\Delta t^*$, which 
varied in the range $10^{-3}-10^{-4}$; typically, a lower $\Delta t^*$ 
is required at high volume fractions or low shear rates. We note that the non-dimensionalization
of the time-step depends on the shear rate ($Pe$). At 
low shear rates ($Pe<1$) , $\Delta t$ is non-dimensionalized by the diffusive time scale,
i.e., $\Delta t^* = D_0/a^2 \Delta t$, while at higher shear rates ($Pe \geq 1$),
$\Delta t$ is non-dimensionalized by the convective time scale, i.e., $\Delta t^* = \dot{\gamma} \Delta t$.
The total non-dimensional integration time $t^*$ was $t^*=500$, with $t$ being non-dimensionalized
in the same way as $\Delta t$. This total integration time was found to be 
sufficient to obtain good statistical averages of various properties
in the apparent steady state -- in most cases, an apparent steady state was evident for
times greater than  $t^* > 150$  as evidenced by various of the suspension properties, like
the orientation distribution function.
The last parameter of interest is the repulsive force prefactor $C_p$, which in the non-dimensional
form is set to $C_p^* =10$.  At low shear rates ($Pe<1$), $C_p$ is non-dimensionalized
as: $C_p^* = C_p/(8\pi\mu a^4/D_0)$, while at higher shear rates ($Pe \geq 1$),
$C_p$ is non-dimensionalization as: $C_p^* = C_p/(8\pi\mu a^2 \dot{\gamma})$. 
The specification of various of the above parameters concludes the formulation section.
We now turn our attention to the results from the numerical simulations, which are presented
in the sections below.

\section{Results and Discussion}

\subsection{Microstructure}\label{sec:struct}
\begin{figure}
\centering
\includegraphics[width=0.7\textwidth]{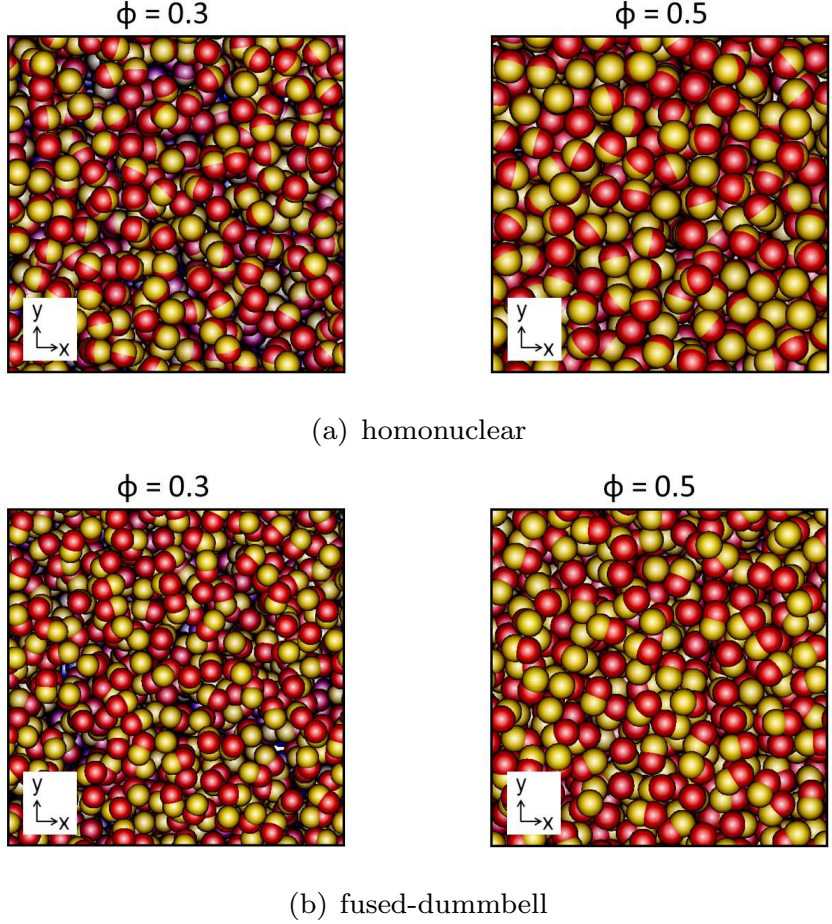}
\caption{Simulation snapshots in the front view ($x-y$ plane) at a shear rate of \pe{10} and 
for two different volume fractions (\ph{0.3} and \ph{0.5}) for (a) homonuclear particles  and 
(b) fused-dumbbell particles.}\label{fig:snap}
\end{figure}

In this section, we present results for the positional (center-of-mass) microstructure in suspensions
of homonuclear and fused-dumbbell particles. Results are presented for volume fractions between \ph{0.3}
and \ph{0.5} and for a wide range of P\'{e}clet numbers varying between \pe{0} and \pe{1000}.
Figure (\ref{fig:snap}) shows some illustrative simulation snapshots at \pe{10} for both 
the homonuclear and the fused-dumbbell particles in \ph{0.3} and \ph{0.5} suspensions.
These snapshots are in the front view, i.e., in the velocity--gradient ($x-y$) plane.
At the lower volume fraction of \ph{0.3}, there are no signs of ordering for 
both the particle shapes. At the higher volume fraction of \ph{0.5}, several regions
with a string like structure aligned in the flow direction are apparent in the homonuclear
particle suspension, though the fused-dumbbell particle suspension appears mostly disordered
even at this volume fraction.

For a more quantitative characterization of the microstructure, we will employ the commonly
used measure of pair distribution function (PDF) denoted $g({x},{y},{z})$. Note that the PDF
gives the conditional probability of finding a particle at a given location
$(x,y,z)$ provided a particle is already present at the origin. Additionally, the PDFs are normalized
such that $g(x,y,z)=1$ in the absence of any correlations between particle positions.
Given the full three dimensional PDF $g({x},{y},{z})$, it is usually instructive
to consider its  projections in various two dimensional planes -- the projection
in the front view ($x-y$ plane) denoted $g(x,y)$ will be of particular interest. Formally, $g(x,y)$ is defined as follows:
\begin{equation}
g({x},{y}) = \frac{1}{2a} \int_{-a}^{+a} g(x,y,z) \, dz.
\end{equation}	

\begin{figure}
\centering
\includegraphics[height=0.8\textheight]{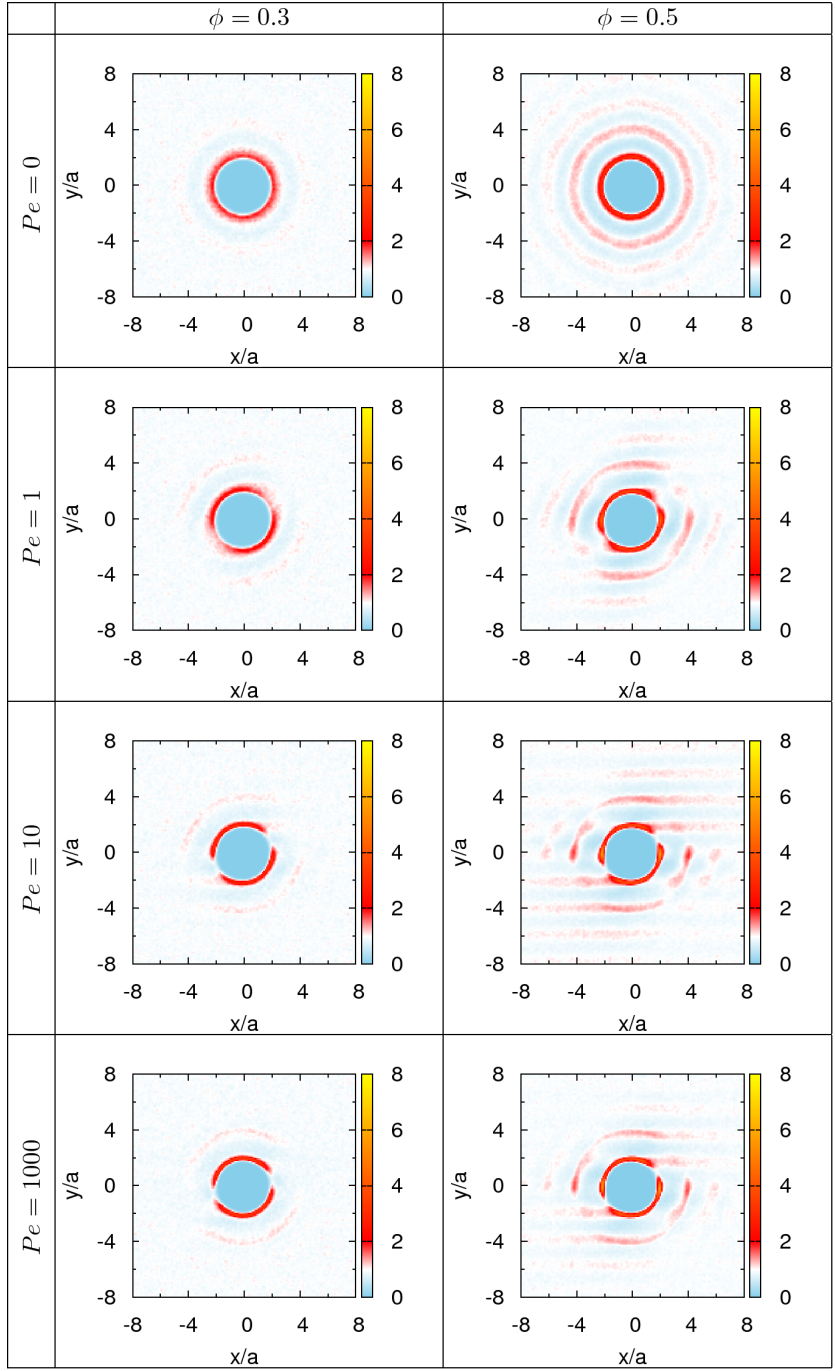}
\caption{Pair distribution function in the front view ($x-y$ plane) for homonuclear particles
at several $Pe$ and two different volume fractions (\ph{0.3} and \ph{0.5}).}\label{fig:hngxy}
\end{figure}

\begin{figure}
\centering
\includegraphics[height=0.8\textheight]{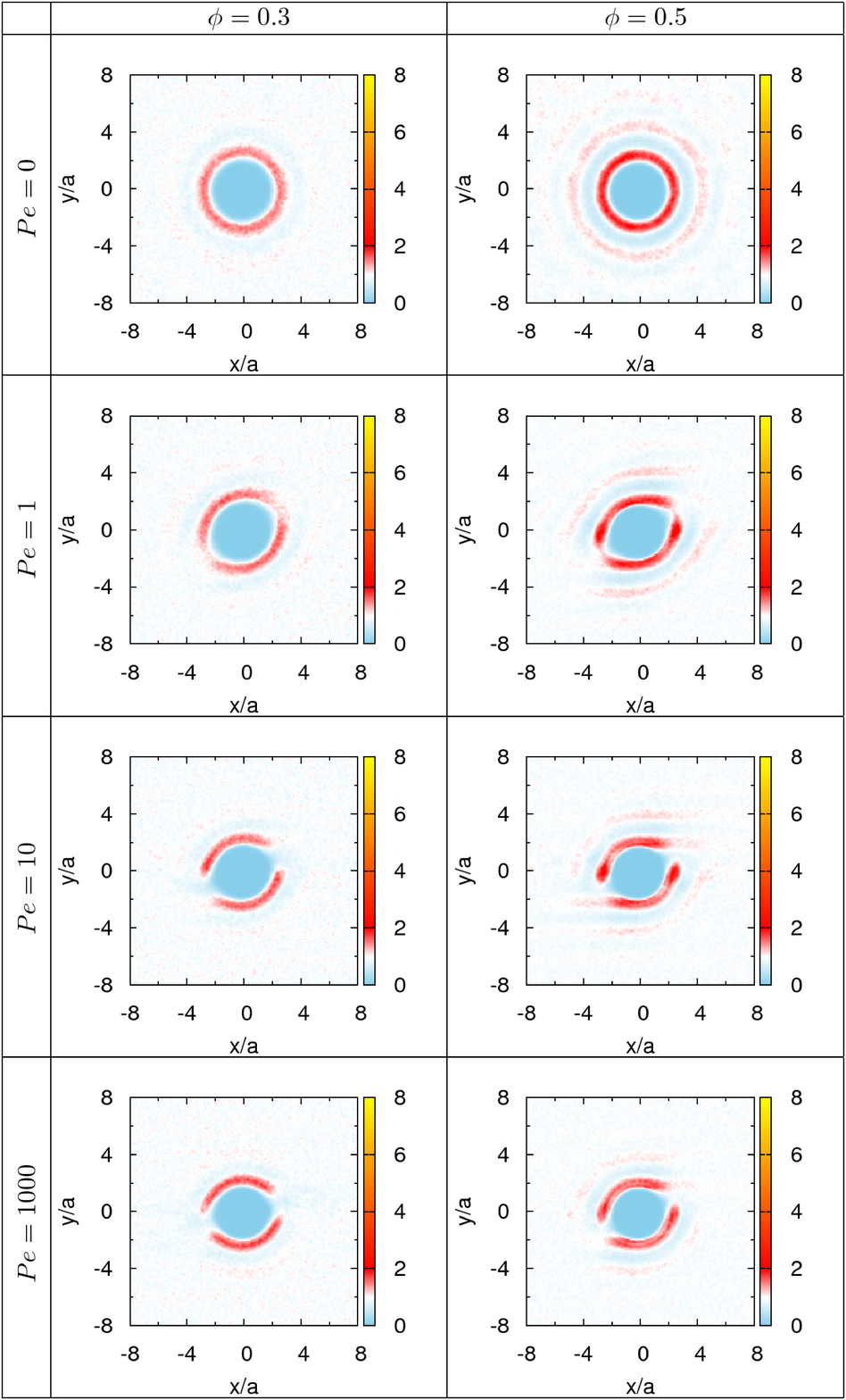}
\caption{Pair distribution function in the front view ($x-y$ plane) for fused-dumbbell particles
at several $Pe$ and two different volume fractions (\ph{0.3} and \ph{0.5}).}\label{fig:fdgxy}
\end{figure}

The PDF in the front view for homonuclear particles in \ph{0.3} and \ph{0.5} suspensions are 
shown in Fig. (\ref{fig:hngxy}) at several P\'{e}clet numbers between \pe{0} and \pe{1000};
figure (\ref{fig:fdgxy}) illustrates the same for fused-dumbbell particle suspensions.
It is appropriate to remark here that the color scheme used in the PDF plot is chosen
to emphasize small differences from the value of one, and, as such, different patterns in these
plots do not necessarily imply magnitudes greatly different from one. Returning to the above
two figures (Figs. \ref{fig:hngxy} and \ref{fig:fdgxy}),
we note that at equilibrium (\pe{0}), a disordered microstructure is apparent
in suspensions of both the homonuclear and fused-dumbbell particles at \ph{0.3};  
in these plots, only the first coordination shell distinctive of an amorphous
fluid like structure is apparent. At the higher volume fraction of \ph{0.5}, the structure stays 
disordered at \pe{0}, though in these cases the second (for fused-dumbbell) or third (for homonuclear)  coordination shells
are also apparent. A disordered structure at equilibrium at these volume fractions is consistent with the 
Monte Carlo simulations of \citet{vega92b}. As per the previous reference, a
suspension of homonuclear particles are in a fluid like state below a volume 
fraction of \ph{0.4945} (called freezing point) and a crystalline solid state above \ph{0.543} (called melting point);
between these two volume fractions the two phases coexist. In fused-dumbbell particle suspensions,
the freezing and melting transitions were found to occur at approximately $\phi=0.6013$ and $\phi=0.6463$.
Typically, in simulations, the ordered phase is observed at volume fractions above the melting transition \citep{speedy97}.
Therefore, a fluid like state is expected at equilibrium in suspensions of
homonuclear as well as fused-dumbbell particles for volume fractions up to \ph{0.5} considered here.
It is interesting to remark here that the ordered phase at the freezing transition
is orientationally disordered (plastic crystal) face centered cubic in suspensions of 
homonuclear particles, while it is orientationally ordered base centered monoclinic in 
fused-dumbbell particle suspensions \citep{vega92b}. It is also pertinent to remark here that a nematic
phase is not expected in dicolloidal particle suspensions. This is due to the fact
that the degree of anisotropy of dicolloids is small, while the nematic phase is stable only
at higher aspect ratios. As an example, for spherocylinders, the nematic phase becomes
stable at an aspect ratio of 4.7 \citep{bolhius97}. 

We next discuss the microstructure in the \ph{0.3} suspensions of homonuclear and fused-dumbell
particles at finite rates of shear. At very low shear rates (\pe{0.01}),
the microstructure is found to be very similar to the equilibrium microstructure
in both the homonuclear and the fused-dumbbell particle suspensions (not shown). As the shear
rate is further increased ($Pe = 1$ in the first column of Figs. \ref{fig:hngxy} and \ref{fig:fdgxy}),
the microstructure becomes distorted -- there is an enhancement in pair probability
in the compressional quadrants, accompanied by its depletion in the extensional quadrants.
This fore-aft asymmetry in the PDF results due to the presence of Brownian forces as discussed
in \citet{brady97} and \citet{foss2000}. Lastly, we note that the microstructure remains disordered
at the highest shear rates investigated here (\pe{1000}) in the \ph{0.3} suspensions
(Figs. \ref{fig:hngxy} and \ref{fig:fdgxy}).

At the higher volume fraction of \ph{0.5}, some interesting features emerge in the PDFs 
of both the particle shapes (Figs. \ref{fig:hngxy} and \ref{fig:fdgxy}). At low shear rates (\pe{0.01}),
the microstructure is similar to the equilibrium microstructure for both the particle shapes (not shown).
However, at \pe{10}, evidence of weak string like ordering is apparent in suspensions of
homonuclear particles (horizontal bands in Fig. \ref{fig:hngxy}), which is sustained 
at higher shear rates, though the ordering become progressively weaker 
(e.g., compare PDF at \pe{10} with PDF at \pe{1000}). In fused-dumbbell particle suspensions, the microstructure
shows similar hints of string like ordering at \pe{10}, though the strength of ordering is considerably weaker
than in suspensions of homonuclear particles (Fig. \ref{fig:fdgxy}).
At further higher shear rates in fused-dumbbell particle suspensions (e.g., \pe{1000}),
the microstructure appears mostly disordered.

The results presented in the current study are consistent 
with previous experimental and computational efforts. In experimental studies on near 
hard sphere suspensions, \citet{ackerson90} noted that 
suspensions which are disordered (liquid-like) at rest stays disordered under shear.
This is consistent with the disordered structure observed in \ph{0.3} suspensions of homonuclear
particles under shear, and also in fused-dumbbell particle suspensions for volume 
fractions up to \ph{0.5} considered here. \citet{ackerson90} also noted that at volume fractions
near the freezing transition, string like structures begin to coexist with amorphous fluid like regions
at intermediate rates of shear -- this is similar to the  weak string like ordering observed in \ph{0.5} homonuclear
particle suspension. 	Similarly, in Stokesian Dynamics simulations on non-Brownian dicolloidal particle suspensions, which is equivalent to  the high $Pe$ limit, \citet{kumar_jor} also observed weak string like ordering  in suspensions
of homonuclear particles at \ph{0.5}. In independent Stokesian dynamics simulations of \citet{brady2002},
again for non-Brownian particles, very strong string like ordering was observed
in suspensions of spheres at \ph{0.52}, which is just above the freezing point of \ph{0.494} in
hard sphere suspensions. It is instructive to note that in suspensions that are
ordered at rest, like in sufficiently charged particle suspensions, very distinct
and rich set of microstructural transitions are observed as a function of shear rate; see, e.g., \citet{chen94,chow95}.
This regime, however, is not relevant to our study here.
Finally, there have been very limited experimental studies on the microstructure of dicolloidal particle 
suspensions under steady shear. The only study we are aware of is that by \citet{mock07},
who studied \textit{charge-stabilized} dicolloidal particle suspensions under shear at a volume fraction of
$\phi=0.42$. The homonuclear particle equivalent investigated in this work was found
to possess a polycrystalline structure at rest, a sliding layer structure at intermediate
rates of shear, and an amorphous structure at further higher shear rates. A direct
comparison of this work with the present effort, though, is not possible due to the presence of 
strong electrostatic repulsion in the former study.

\subsection{Orientation behavior}\label{sec:oc}

\begin{figure}
\centering
\includegraphics[width=0.7\textwidth]{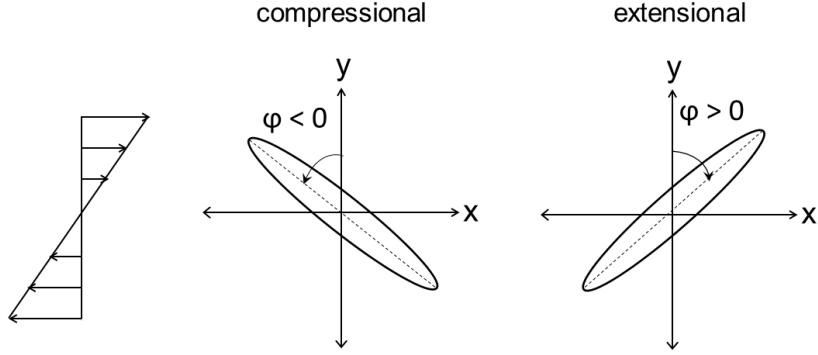}
\caption{Examples of particles aligned in the compressional and the extensional quadrants.}\label{fig:ec}
\end{figure}

We next present the results on the orientation behavior in homonuclear and
fused-dumbbell particle suspensions as a function of $Pe$ for volume fractions between
\ph{0.3} and \ph{0.5}. The statistical distribution of the particle orientation will be characterized
by the orientation distribution function (ODF) denoted $P(\theta,\varphi)$, where
$\theta$ and $\varphi$ are spherical coordinates as illustrated in Fig.~(\ref{fig:defn}); also see 
Fig.~(\ref{fig:ec}) for illustrations of the angle $\varphi$.
In words, $\theta$ is the polar angle between the director \nvec of the particle and the z-axis,
while $\varphi$ is the azimuthal angle between its projection in the $x-y$
plane and the $y$-axis measured clockwise. Based on the ODF $P(\theta,\varphi)$, the differential 
probability  of finding a particle with an orientation
between ($\theta$, $\theta$+$d\theta$) and ($\varphi$,$\varphi$+$d\varphi$), $dN$, can be expressed as follows:
\begin{equation}
dN = \frac{1}{4\pi}\,P(\theta,\varphi) \, \sin \theta  \, d\theta  \, d\varphi.
\end{equation}
Additionally, we note that the $P(\theta,\varphi)$ satisfies the following normalization constraint:
\begin{equation}
\int_{-\pi/2}^{\pi/2} \, \int_{0}^{\pi/2} \, P(\theta,\varphi) \, \sin \theta  \, d\theta  \, d\varphi  = 1,
\end{equation}
where we have restricted the limits of integration in $\theta$ to $[0,\pi/2]$ and in $\varphi$ to 
$[-\pi/2,\pi/2]$, which is half of the corresponding overall ranges; this is 
sufficient here owing to the symmetry of the ODF $P(\theta,\varphi)$. We next introduce
one dimensional ODFs, $P(\theta)$ and $P(\varphi)$, which are defined as follows:
\begin{subequations}
\begin{equation}
P(\theta) = \int_{-\pi/2}^{\pi/2} P(\theta, \varphi)\, d \varphi,
\end{equation}
\begin{equation}
P(\varphi) =  \pi  \int_{0}^{\pi/2} P(\theta, \varphi) \, \sin \theta \, d \theta.
\end{equation}
\end{subequations}
We will use $P(\theta)$ and $P(\varphi)$ extensively in lieu of the full ODF $P(\theta,\varphi)$.
It is important to emphasize here that the definitions of $P(\theta)$ and $P(\varphi)$ are such that a uniform 
distribution in the orientation space will result in $P(\theta)=1$ and $P(\varphi)=1$. 
For later use, we also define a suspension averaged $\theta$ and $\varphi$, denoted 
$\theta_m$ and $\varphi_m$, as follows: 
\begin{subequations}
\begin{equation}
\theta_m = \int_{0}^{\pi/2} \theta \, P(\theta)  \, \sin \theta \, d \theta, 
\end{equation}
\begin{equation}
\varphi_m = \frac{1}{\pi} \int_{-\pi/2}^{\pi/2} |\varphi| \, P(\varphi)  \, d \varphi.
\end{equation}
\end{subequations}
Note the absolute value term `$|\varphi|$' in the definition of $\varphi_m$ above.
A simple interpretation of $\theta_m$ is that it gives the mean inclination of the
particle director with the $z$ axis. Similarly, $\varphi_m$ gives the mean inclination
of the projection of the particle director in the $x-y$ plane with the $y$ axis.
As should be obvious from the above definitions, $\theta_m$ and $\varphi_m$ have the 
following bounds: $0 \leq \theta_m \leq \pi/2$ and $0 \leq \varphi_m \leq \pi/2$. 
In a suspension with a completely random orientation distribution, these two averages
will be denoted by $\theta_m^0$ and $\varphi_m^0$. We then define $\Delta \theta_m$ and $\Delta \varphi_m$ for 
an arbitrary suspension as follows:
\begin{subequations}
\begin{equation}
\Delta \theta_m = \theta_m - \theta_m^0,
\end{equation}
\begin{equation}
\Delta \varphi_m = \varphi_m - \varphi_m^0.
\end{equation}
\end{subequations}
When $\Delta \theta_m$ is negative, it will imply an enhanced vorticity alignment in comparison
to an isotropic distribution, while positive $\Delta \theta_m$ will imply an enhanced
alignment in the velocity--gradient plane (i.e., away from the vorticity axis) 
in comparison to an isotropic distribution. 
A positive $\Delta \varphi_m$ will imply an enhanced alignment with the velocity axis in
the velocity--gradient plane, while a negative $\Delta \varphi_m$ will imply an enhanced alignment 
with the gradient axis in the velocity--gradient plane, both, of course, are relative to 
an isotropic distribution.

In addition to the above measures, it will be useful to employ direct measures 
of the particle orientation relative to the three coordinate axes $x$, $y$, and $z$. 
In the present work, this measure is provided by the orientation order parameter
$S_{ii}$ defined relative to the axis $i$, where $i$ could be $x$, $y$, or $z$.
More specifically,  the orientational order parameter $S_{ii}$ is defined as the 
average over all particles of 
\begin{equation}\label{eq:Sii}
S_{ii} = \frac{3}{2}n_i n_i - \frac{1}{2}.
\end{equation}
Some representative values of $S_{ii}$ are as follows. 
For a random orientation distribution $S_{ii}=0$; for perfect alignment of particles with axis
$i$, $S_{ii}=1.0$; and for particles aligned in the plane perpendicular to axis $i$,
$S_{ii}=-0.5$. It is evident from the above definition that an increased degree of particle alignment 
with an axis $i$ will result in a higher value of $S_{ii}$.

Having introduced the various measures for characterizing the orientation behavior, we turn 
our attention to the results for these measures.
We will first present the orientation distribution in the $\theta$ coordinate as a function of $Pe$, 
volume fraction and particle shapes in Sec.~(\ref{sec:tht}). Following this, we will present the
orientation distribution in $\varphi$ coordinate in Sec.~(\ref{sec:vp}). Lastly, we will present 
the results for the orientation order parameters in Sec.~(\ref{sec:sii}).

\begin{figure}[!tb]
\centering
\includegraphics[width=0.9\textwidth]{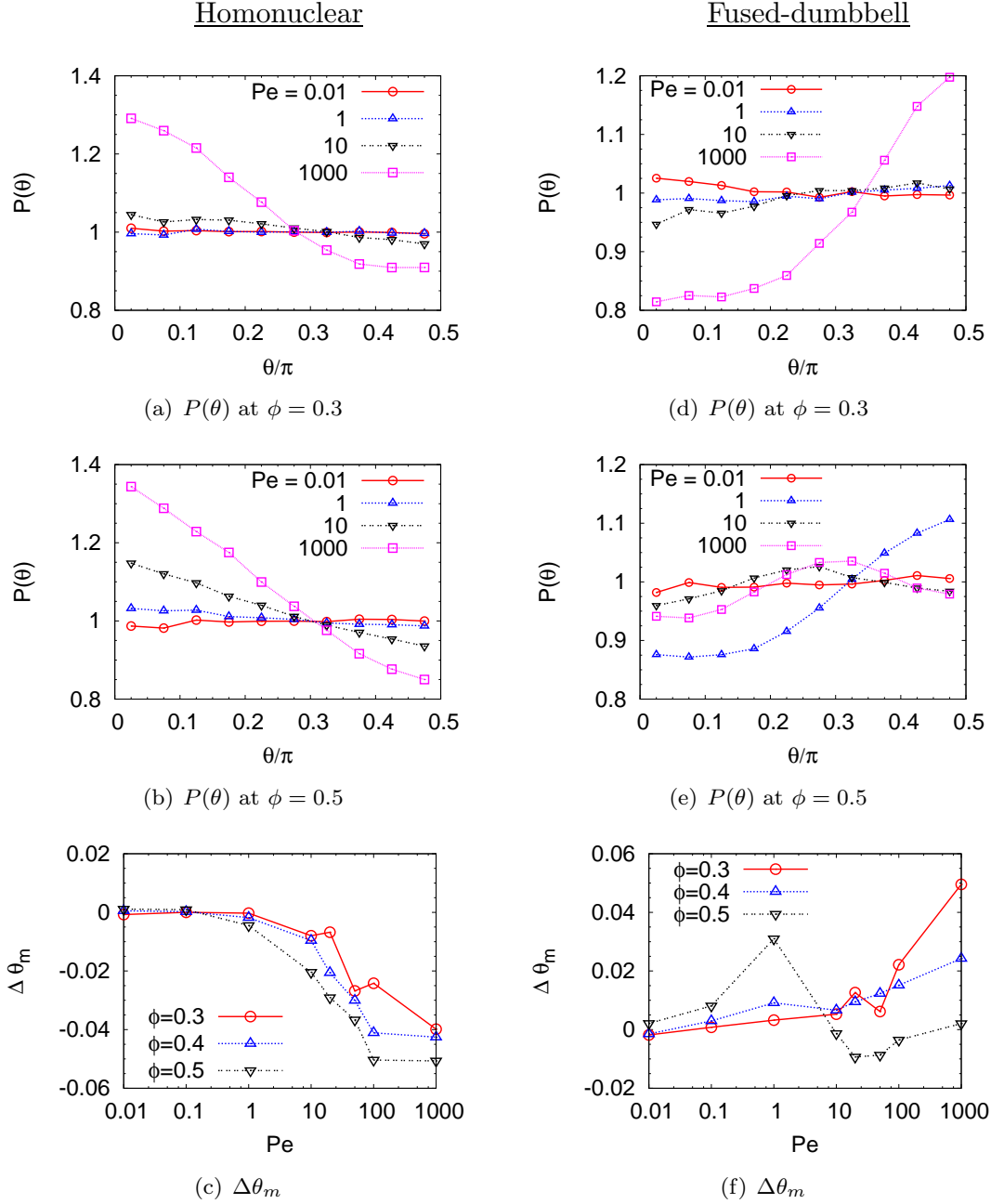}
\caption{$P(\theta)$ and $\Delta \theta_m$ for (a), (b), and (c) homonuclear
particles and (d), (e) and (f) fused-dumbbell particles.}\label{fig:tht}
\end{figure}

\subsubsection{Orientation distribution in $\theta$ coordinate}\label{sec:tht}
Figures (\ref{fig:tht}a) and (\ref{fig:tht}b) show the orientation distribution function $P(\theta)$ 
at several P\`{e}clet numbers in homonuclear particle suspensions with \ph{0.3} and \ph{0.5}, respectively.
The same two plots for suspensions of fused-dumbbell particles are shown
in Figs.~(\ref{fig:tht}d) and (\ref{fig:tht}e), respectively. The change in the suspension 
averaged $\theta$ over an isotropic distribution, $\Delta \theta_m$, is shown in Fig.~(\ref{fig:tht}c)
for homonuclear particle suspensions, while it is shown in Fig.~(\ref{fig:tht}f)
for fused-dumbbell particle suspensions.

We first consider the orientation behavior in suspensions of homonuclear particles.
At low shear rates (\pe{0.01}) in the \ph{0.3} suspension, $P(\theta)$ has a value
close to one, implying a random distribution in $\theta$ at these shear rates (Fig.~\ref{fig:tht}a).
As the shear rate is further increased, an enhancement in probability at low $\theta$ values
is observed, indicating an increased vorticity alignment; this enhancement is particularly
strong at $Pe=1000$ (Fig.~\ref{fig:tht}a).
We see a very similar trend at the higher volume fraction of \ph{0.5} (Fig.~\ref{fig:tht}b),
namely, a random distribution in $\theta$ at low shear rates (\pe{0.01}), which shifts towards 
lower $\theta$ values, i.e., increased vorticity alignment, at higher shear rates (e.g. \pe{1000}).
These trends in the orientation behavior with shear rate are also evident in the plot of $\Delta \theta_m$ in Fig.~(\ref{fig:tht}c).
At low shear rates, $\Delta \theta_m$ is nearly zero indicating a random orientation distribution,
while it becomes progressively more negative with increasing shear rates, implying an increased vorticity
alignment at higher shear rates. It is also evident in the plot that $\Delta \theta_m$ is
more negative at higher volume fractions, implying a greater vorticity
alignment at higher volume fractions in homonuclear particle suspensions.

We next discuss the orientation results in suspensions of fused-dumbbell particles.
Focusing on the plot of $P(\theta)$ in \ph{0.3} fused-dumbbell particle 
suspension (Fig.~\ref{fig:tht}d), it is obvious that the distribution at low shear rates (\pe{0.01}) is
close to random. With increasing shear rates, however, $P(\theta)$ changes from a random
distribution to a distribution skewed towards higher $\theta$ values, indicating an enhanced alignment
in the velocity--gradient plane (Fig.~\ref{fig:tht}d). This trend is clearly opposite to the 
trend observed in homonuclear particle suspensions. At the higher volume fraction (\ph{0.5}), 
the behavior is more complex (Fig.~\ref{fig:tht}e). In this case, at intermediate shear rates (\pe{1}),
an enhanced alignment in the velocity--gradient plane (i.e., at $\theta$ values) is observed.
With further increase in shear rate, however, the enhanced particle alignment in the velocity--gradient
plane is attenuated (e.g., at \pe{1000}), and the distribution gradually shifts towards vorticity
alignment, though it is not as complete or as strong as in homonuclear particle suspensions.
We confirm these observations by examining the plot of $\Delta \theta_m$  in Fig.~(\ref{fig:tht}f).
It is evident from the figure that, at low volume fractions (\ph{0.3}), $\Delta \theta_m$ increases nearly monotonically with increasing
shear rate, indicating an increasing alignment in the velocity--gradient plane. However, at \ph{0.5}, $\Delta \theta_m$ 
initially increases with increasing shear rate till \pe{1}, and thereafter it decreases 
and becomes negative with increasing shear rates before settling at a near zero value.

In a recent study \citep{kumar_jor}, we investigated the orientation distribution
in the $\theta$ coordinate in sheared non-Brownian suspensions of dicolloids.
More precisely, the orientation distribution into Jeffery's orbits 
was investigated, which approximately corresponds to a distribution 
in the $\theta$ coordinate for small aspect ratio particles like dicolloids.
This study indicates that in low volume fraction suspensions,
both the homonuclear and fused-dumbbell particles exhibit increased particle alignment
in the velocity-gradient plane. The increase in the velocity--gradient 
alignment in suspensions of fused-dumbbell particles was found to be considerably higher than in the 
suspensions of homonuclear particles. At higher volume fractions, both the particle 
shapes were found to exhibit a shift towards enhanced vorticity alignment, though
the degree of vorticity alignment was much weaker in suspensions of fused-dumbbells
than in suspensions of homonuclear particles. The origins of the contrasting  behavior at low and high volume
fractions was attributed to two different mechanisms at play in these two parameter regimes.
At low volume fractions, on an average, it was observed that a particle aligned with the vorticity axis
underwent comparatively stronger fluctuation in its orientation during the course of a collision
with another particle. This behavior leads to non-uniform fluctuations in the orientation space 
with a stronger fluctuation observed for vorticity aligned particles.
Additionally, the non-uniformity of fluctuations was found to be stronger and distributed over
a wider region in the orientation space for the more anisotropic fused-dumbbell particles
than the homonuclear particles. As discussed in \citet{koch88}, these non-uniform fluctuations
will result in a drift away from regions of higher fluctuations towards regions of lower fluctuations.
Predictions based on this mechanism easily resolves two of the key observations at low volume fractions,
namely, an increased alignment in the velocity--gradient plane
in both the homonuclear and the fused-dumbbell particle suspensions, and a
comparatively greater alignment in the velocity--gradient plane in fused-dumbbell particle suspensions. 
At higher volume fractions, the fluctuations were found to be nearly uniform 
in the orientation space and hence cannot lead to a non-uniform orientation
distribution. The high volume fraction behavior was instead found to be driven by the 
negative second normal stress difference exhibited by these suspensions at high volume fractions
(note that at low volume fractions, the normal stress differences are very weak). A detailed
discussion of this mechanism is beyond the scope of the current work and the interested reader
is referred to the original publication \citep{kumar_jor}. The greater vorticity alignment
of the homonuclear particles than that of the fused-dumbbell particles at high volume fractions
was shown to result from the lower degree of anisotropy of homonuclear particles. In essence this 
implies that the homonuclear particles have a greater rotational ``mobility" than fused-dumbbell particles in
 concentrated suspensions, and hence the second normal stress difference is a more effective driving force
in the former case.

With the aid of the above discussion, we can provide some mechanistic 
insights into the observed orientation behavior in the current study.
Firstly, in all cases, a random orientation at low shear rates is expected due to the dominant effect of the 
Brownian forces in this regime. The high shear rate regime is well 
explained by the non-Brownian limit discussed above. The non-Brownian limit
consistently explains the enhanced vorticity alignment observed 
in suspensions of homonuclear particles at high shear rates and 
at all volume fractions considered here.
It also consistently explains the enhanced velocity--gradient plane alignment 
in low volume fraction suspensions of fused-dumbbell particles, 
and the (weak) vorticity alignment in higher volume fraction suspensions of 
fused-dumbbell particles, both in the high shear rate limit. 
The only regime that remains to be resolved is the intermediate shear rate
regime. Intuitively, one may expect that the behavior at intermediate shear 
rates will be ``intermediate" of the high and the low shear rate behaviors.
This is certainly found to be true, but there is one exception --
 the intermediate shear rate behavior in the 
 \ph{0.5} fused-dumbbell particle suspension (Fig.~\ref{fig:tht}f).
In this case a random distribution is observed at low 
shear rates, while a weak vorticity alignment is observed at higher
shear rate. The intermediate shear rate, however, exhibits enhanced
velocity-gradient alignment. The origins of this behavior is not entirely clear, 
though it is possible that the ``non-uniform fluctuation"
mechanism plays a role at intermediate shear rates in high volume fraction suspensions as well (at very low 
shear rates Brownian motion will dominate any other effect). Further studies
are clearly necessary to confirm this mechanism.

\begin{figure}[!tb]
\centering
\includegraphics[width=0.9\textwidth]{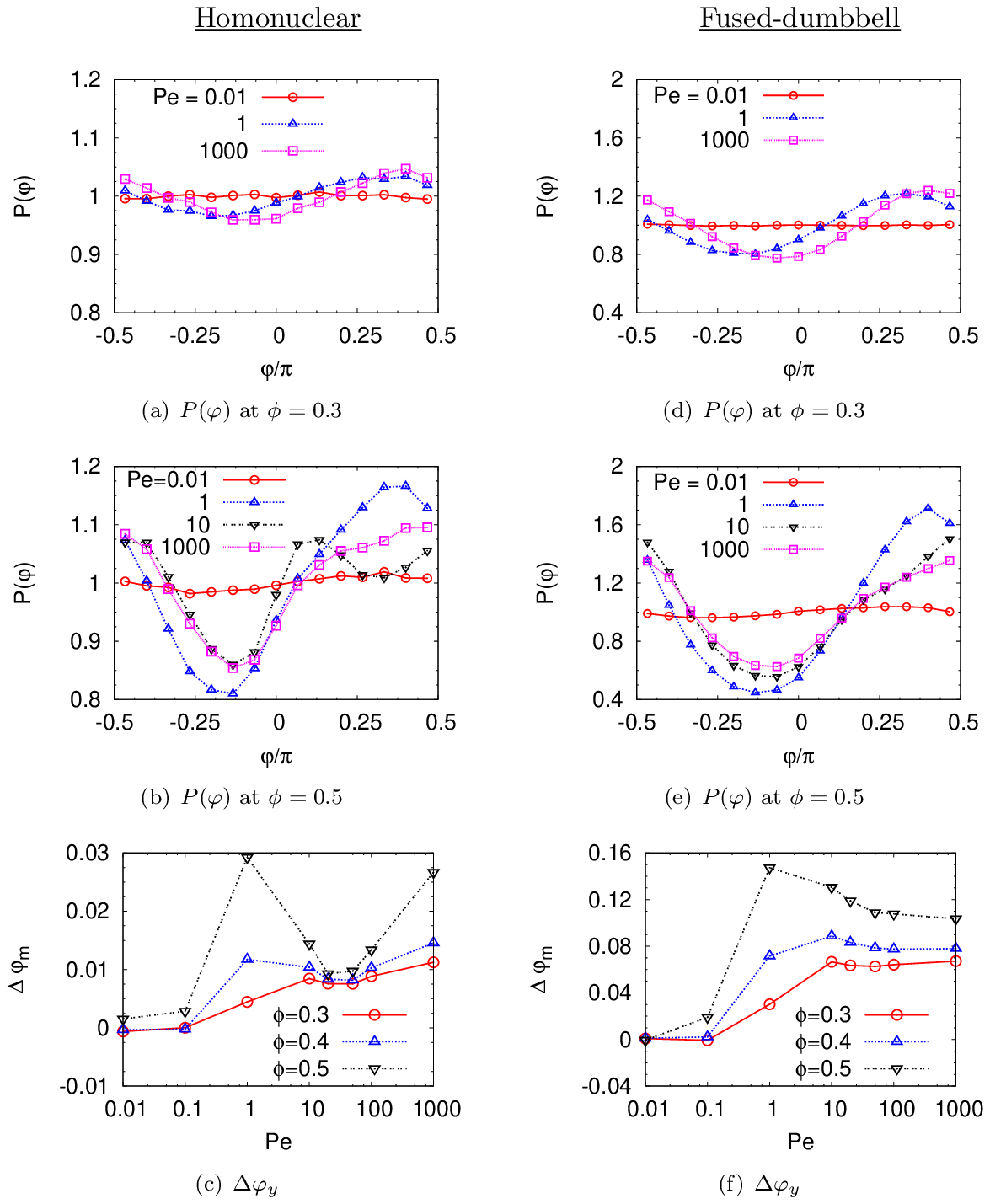}
\caption{$P(\varphi)$ and $\Delta \varphi_m$ for (a), (b), and (c) homonuclear
particles and (d), (e) and (f) fused-dumbbell particles.}\label{fig:vp}
\end{figure}

\subsubsection{Orientation distribution in $\varphi$ coordinate}\label{sec:vp}
The orientation distribution in the $\varphi$ coordinate will be characterized 
by the ODF $P(\varphi)$ as well as by $\Delta \varphi_m$, the change in the suspension averaged $\varphi$
in comparison to an isotropic distribution. For homonuclear particles, 
Fig.~(\ref{fig:vp}a) shows $P(\varphi)$ in \ph{0.3} suspension  and Fig.~(\ref{fig:vp}b) shows the same in \ph{0.5} suspension;
the plot for $\Delta \varphi_m$ is shown in Fig.~(\ref{fig:vp}c). For fused-dumbbell particles, Fig.~(\ref{fig:vp}d) shows $P(\varphi)$ 
in \ph{0.3} suspension  and Fig.~(\ref{fig:vp}e) shows the same in \ph{0.5} suspension;
the plot for $\Delta \varphi_m$ is shown in Fig.~(\ref{fig:vp}f).

We begin our discussion by examining the ODF $P(\varphi)$ in homonuclear particle suspensions 
at \ph{0.3} (Fig.~\ref{fig:vp}a). At low shear rates (\pe{0.01}), 
a random distribution is obvious as $P(\varphi) \approx 1$.
At intermediate rates of shear (\pe{1}) for the same suspension, we find a decrease 
in probability corresponding to orientations in the compressional quadrant ($-0.5<\varphi/\pi<0$) 
accompanied by an increase in probability corresponding to orientations in the extensional quadrant ($-0.5<\varphi/\pi<0$).
Examples of particles aligned in the compressional and extensional quadrants are illustrated in Fig.~(\ref{fig:ec}).
At higher shear rates (e.g., \pe{1000} in Fig.~\ref{fig:vp}a), the height of probability maxima
as well as the depth of probability minima increases. Additionally, the probability minima is found to shift
closer to the gradient axis ($\varphi = 0$) and the probability maxima closer to the velocity axis ($\varphi=\pi/2$)
with increasing shear rates (e.g., compare \pe{1} and \pe{1000} in Fig.~\ref{fig:vp}a).
Turning our attention to suspensions of fused-dumbbell particles at the same volume fraction (\ph{0.3}), 
we observe the same trends in the ODF $P(\varphi)$ as in the suspension of homonuclear particles (compare Fig. \ref{fig:vp}a and
\ref{fig:vp}d).

At the higher volume fraction of \ph{0.5}, the ODF $P(\varphi)$ shows additional 
features in both the homonuclear and the fused-dumbbell particle suspensions (Figs.~\ref{fig:vp}b and \ref{fig:vp}e).
In the homonuclear particle suspension, an expected random distribution 
is evident at \pe{0.01} (Fig.~\ref{fig:vp}b). At intermediate shear rates (\pe{1}),
similar to the \ph{0.3} suspension, an enhancement in probability corresponding to
orientations in the extensional quadrant and a reduction in probability corresponding
to orientations in the compressional quadrant is observed (Fig.~\ref{fig:vp}b). However,
with further increase in shear rate (\pe{10}), the height of the probability maxima as well as the depth of 
the probability minima decreases (Fig.~\ref{fig:vp}b). This is in contrast to the behavior at low volume fractions
discussed above. This trend at \pe{10} is also evident in fused-dumbbell particle suspensions at this
volume fraction (Fig. \ref{fig:vp}e).
At further higher shear rates (e.g., \pe{1000}), the height of the maxima as well as the depth of the minima 
generally increases in suspensions of the homonuclear particles (Fig.~\ref{fig:vp}b), but 
decreases slightly in suspensions of fused-dumbbell particles (Fig.~\ref{fig:vp}e).

The trends in the ODF $P(\varphi)$ discussed above are confirmed in the $\Delta \varphi_m$ plots shown in Fig.~(\ref{fig:vp}c)
for homonuclear particles and in Fig.~(\ref{fig:vp}f) for fused-dumbbell particles.
For both homonuclear and fused-dumbbell particles, there is a general tendency of increasing $\Delta \varphi_m$ with 
increasing volume fractions. Further, for both particle shapes, $\Delta \varphi_m$ increases with increasing 
shear rates at low volume fractions (e.g., \ph{0.3}). At high volume fractions, however, 
$\Delta \varphi_m$ does not necessarily increase with increasing shear rates. This is particularly 
evident in the \ph{0.5} homonuclear particle suspensions, where $\Delta \varphi_m$ is found to decrease
in a small shear rate window around \pe{10} (Fig. \ref{fig:vp}c).
In fused-dumbbell suspensions at high volume fractions, $\Delta \varphi_m$ is found to decrease 
at all shear rates above $Pe > 1$ (Fig. \ref{fig:vp}f).
Another feature apparent in the plots of $\Delta \varphi_m$ for the two particle shapes is
that its magnitude is considerably larger in suspensions of fused-dumbell particles
than in suspensions of homonuclear particles. It is worth recalling at this point
that $\Delta \varphi_m$ has important implications for results on the flow and the gradient alignment of the particle,
e.g., a higher positive $\Delta \varphi_m$ will imply a greater degree of flow alignment at the expense of gradient 
alignment in the velocity--gradient plane.  These aspects are discussed further in Sec. (\ref{sec:sii}).

\begin{figure}[!tb]
\centering
\includegraphics[width=0.4\textwidth]{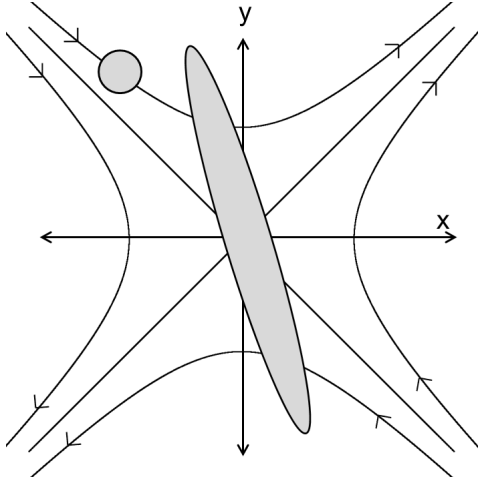}
\caption{A cartoon showing a slender body subjected to an extensional flow. Also shown 
	is a small sphere ``colliding" with the slender body.}\label{fig:ec1}
\end{figure}

Some of the trends in the ODF $P(\varphi)$ as a function of shear rate have striking similarities
with the corresponding trends in suspensions of Brownian axisymmetric particles at infinite dilution.
We discussed the orientation behavior in such systems in the introduction and we summarize the
key observations here. At equilibrium, the orientation distribution is expectedly isotropic. 
As the suspension is sheared at very low shear rates, to leading order, the vorticity component of the 
shear flow has no effect on the orientation distribution due to the isotropic 
distribution at equilibrium. The effect of the extensional component of the flow 
is to rotate the particle away from the compressional axis towards an alignment with 
the extensional axis. As a consequence, the probability near the compressional axis
is reduced, while the probability near the extensional axis is enhanced \citep{brenner74}.
The higher order approximations result from the vorticity as well as the 
extensional component of the flow \citep{brenner74}. In particular, the effect 
of the vorticity component of the flow is to rotate the distorted orientation distribution
clockwise, such the probability maxima in the ODF shifts closer to the velocity axis, while
the probability minima shifts closer to the gradient axis.

It is evident that the ODF $P(\varphi)$ in the present effort exhibits some trends similar
to that in suspensions at infinite dilution. However, the mechanism of the distortion
of the ODF upon the onset of shear is very different in the two cases. 
First of all, we remind the reader that the effective single particle resistance
tensor $\mathbf{R}_0$ of dicolloids in FLD is diagonal, which is qualitatively
similar to the resistance tensor of an isolated sphere. Hence a single dicolloid
by itself cannot exhibit a preferred alignment with any axis upon being subjected to shear.
So, the distortion of the ODF in the present case clearly results from particle-particle interactions,
which are expected to dominate anyway in concentrated suspensions.
The precise mechanism of this distortion is easily understood with the help of a
highly idealized example presented below.

Consider a single slender axisymmetric particle subjected to extensional flow,
as shown in Fig. (\ref{fig:ec1}). This single particle by itself
cannot rotate upon being subjected to extensional flow due to reasons discussed above.
Now consider a second particle, which for simplicity is taken as a relatively small sphere
in Fig. (\ref{fig:ec1}). The sphere, to first approximation, will move along the streamlines
and consequently ``collide" with the slender body causing it to rotate clockwise
towards the extensional axis. It is easy to  see that this will be case for
most, if not all, starting positions of the second particle.
Thus, in this idealized case, we see that the effect of particle collisions  on the 
orientation dynamics of the slender body is much the same as that of slender body
by itself in the extensional flow. By analogy of the orientation behavior at infinite dilution in \citet{brenner74}, 
we may expect an enhanced orientation probability along the extensional
axis and a reduced orientation probability along
the compressional axis at low shear rates in the present case as well. This is obviously
the case here (see Fig. \ref{fig:vp}).
Furthermore, at higher shear rates ($Pe$), the distorted ODF is expected to rotate clockwise so as to 
shift the probability maxima closer to the velocity axis and the probability 
minima closer to the gradient axis. This again is very similar to the observations
in the present effort. However, there are some additional features in the ODF $P(\varphi)$ not resolved
yet. At \pe{10} in the higher volume fraction suspensions of both the 
homonuclear and the fused-dumbbell particles, the depth of the minima
as well as the height of the maxima of $P(\varphi)$ decreases from the corresponding
values at \pe{1}; see Figs. \ref{fig:vp}(b) and \ref{fig:vp}(e). 
It appears that this behavior results due to the formation of weak string 
like ordering around \pe{10}, particularly in suspensions of homonuclear particles.
An obvious consequence of the presence of string like ordering will be to reduce the 
collision frequency between the particles, which, as discussed above, is a key driver
of the distortion in the ODF. 
Yet another feature of the ODF $P(\varphi)$ in fused-dumbbell particle suspensions at
high volume fractions and high shear rates is that it continues to display a decrease in the height
of probability maxima and the depth of probability minima with increasing 
$Pe$  (Fig. \ref{fig:vp}e). This cannot result from a string like ordering transition,
as the microstructure is completely disordered in these suspensions at high shear rates (see Sec. \ref{sec:struct}). 
A likely origin of this behavior could be the hydrocluster mechanism proposed by \citet{wagner06} -- at high shear rates, 
large clusters of particles are typically formed, and these clusters of particles
are expected to have a lower degree of anisotropy than the individual particles themselves. 
It is likely that the formation of more spherical clusters will result in a reduced
distortion of the ODF. Lastly, we note that the trends in $\Delta \varphi_m$ follows directly 
from the corresponding trends in the ODF $P(\varphi)$, and hence will not be
explicitly discussed here. 

\begin{figure}[!tb]
\centering
\includegraphics[width=0.9\textwidth]{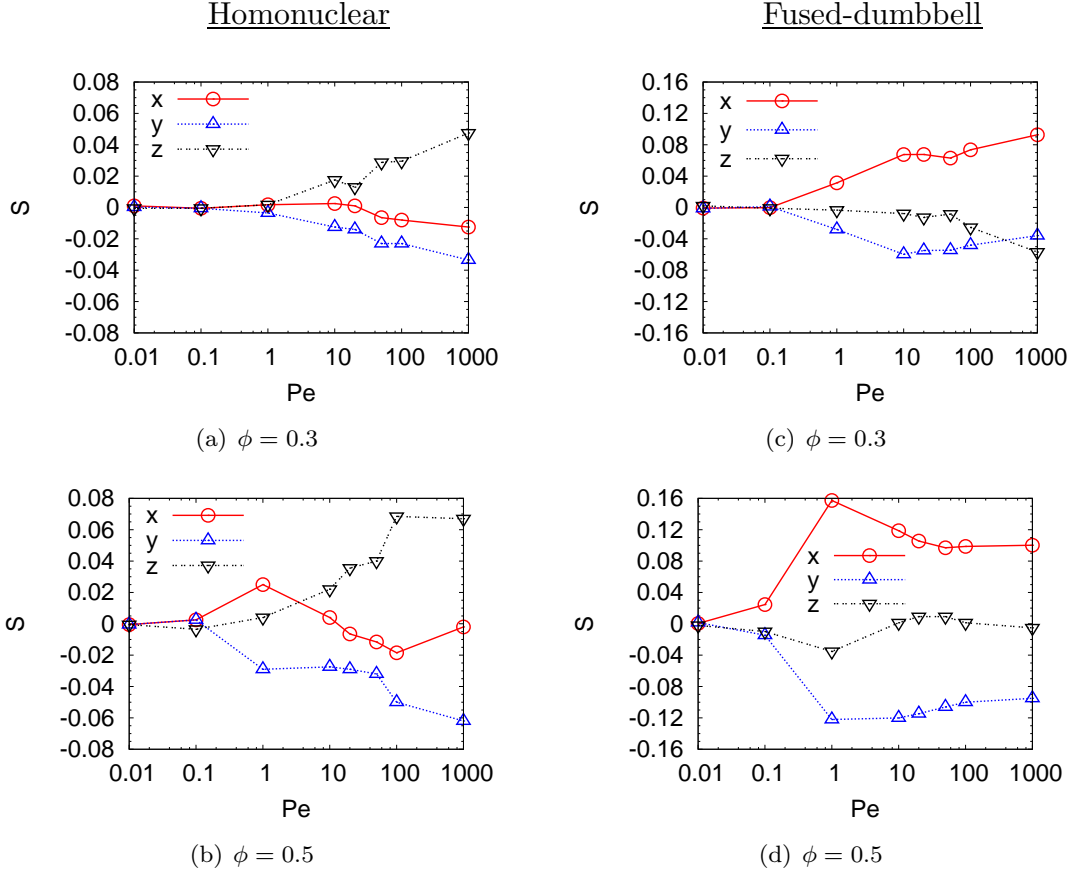}
\caption{Orientation order parameters as a function of shear rate for (a) and (b) homonuclear particles and (c) and (d) 
fused-dumbbell particles at two different volume fractions (\ph{0.3} and \ph{0.5}).}\label{fig:sii}
\end{figure}

\subsubsection{Orientation order parameters}\label{sec:sii}
We discussed above the orientation distribution functions $P(\theta)$ and $P(\varphi)$.
These measures are particularly useful for gaining a 
mechanistic understanding of the orientation behavior. However, 
a more natural way of presenting the orientation results is
to quantify the particle alignment with the three coordinate 
axes: velocity (x), gradient (y), and vorticity (z).
With this goal, we present the results for the  orientation
order parameters $S_{xx}$, $S_{yy}$, and $S_{zz}$ below.

Figures \ref{fig:sii}(a) and  \ref{fig:sii}(b) respectively show the orientation
order parameters in \ph{0.3} and \ph{0.5} suspensions of homonuclear particles as a function of $Pe$.
In \ph{0.3} suspensions, at \pe{0.01}, all three orientation order
parameters are nearly zero, consistent with a random orientation at this
low shear rate. With increasing shear rates, $S_{zz}$ is found to increase,
indicating an increased vorticity alignment with increasing shear rate as discussed
in Sec. \ref{sec:tht}. In contrast, $S_{xx}$ as well as $S_{yy}$ decreases with increasing shear rate;
the decrease in $S_{xx}$ is smaller than the decrease in $S_{yy}$. The decrease in $S_{xx}$ 
and $S_{yy}$ primarily result due to the increase in $S_{zz}$, which leaves a smaller 
fraction to be distributed between $S_{xx}$ and $S_{yy}$. The fact that $S_{xx}$ decreases
slower than $S_{yy}$ is a direct consequence of the orientation distribution in the
$\varphi$ coordinate -- with increasing shear rate, flow alignment is favored at the 
expense of gradient alignment as discussed in Sec. \ref{sec:vp}. At the
higher volume fraction of \ph{0.5}, the trends in orientation order parameters 
are very similar to the trends observed in \ph{0.3} suspension -- the order
parameter $S_{zz}$ increases with shear rate, while $S_{yy}$ decreases with 
shear rate. The only difference is that, at \ph{0.5},  $S_{xx}$ does not decrease monotonically
with increasing shear rate as is the case in the \ph{0.3} suspension. As noted above, 
the instances in which $S_{xx}$ increases with shear rate results due to the enhanced 
flow alignment of the particle in the velocity-gradient plane at the expense of the gradient 
alignment; see Sec. \ref{sec:vp}.
 
We next consider the orientation order parameters for fused-dumbbell particles 
in \ph{0.3} and \ph{0.5} suspensions, which are shown in Figs. \ref{fig:sii}(c) and \ref{fig:sii}(d), 
respectively. In the \ph{0.3} suspension, the order parameter $S_{zz}$ decreases with 
increasing shear rate, which is the direct consequence of increasing $\Delta \theta_m$ with increasing
shear rate; see Sec. \ref{sec:tht}. The order parameter $S_{xx}$ monotonically increases with
increasing $Pe$. This results both due to the monotonically increasing $\Delta \theta_m$ with increasing $Pe$, implying an
increasing velocity--gradient alignment with increasing $Pe$, as well as a (nearly) monotonically
increasing $\Delta \varphi_m$ with increasing shear rate, indicating an increasing flow alignment in the velocity--gradient plane.
The order parameter $S_{yy}$ initially decreases with increasing shear rate, which primarily results due to the 
increase in $S_{xx}$. However, with further increase in shear rate, $S_{yy}$ is found to increase, which can be
attributed to the orientation drift away from the vorticity axis at higher shear rates,
essentially leaving a greater fraction to be distributed in the flow and the gradient direction.
At the higher volume fraction of \ph{0.5}, the order parameter $S_{zz}$, consistent with the trends in $\Delta \theta_m$, 
initially decreases with increasing shear rates, but is then found to increase for $Pe > 1$ and
subsequently levels off at near zero values at very high shear rates. The order parameter  $S_{xx}$ shows a non-monotonic
behavior with increasing shear rate: it initially increases with increasing shear rate reaching a maxima at \pe{1};
thereafter, it decreases with further increase in shear rate and eventually levels off at the highest 
shear rates considered here.
In conjuction with the non-monotonic $S_{xx}$, the variation in the order parameter $S_{yy}$ with increasing 
shear rate is also non-monotonic, though in the opposite direction: $S_{yy}$ initially 
decreases with increasing shear rate, and thereafter it increases before it levels off 
at the high shear rate plateau. 
The non-monotonic variation in $S_{xx}$ (or $S_{yy}$) with shear rate primarily results from the corresponding
non-monotonic variation of $\Delta \varphi_m$ with shear rate. As discussed in Sec. (\ref{sec:vp}), 
the non-monotonic variation in $\Delta \varphi_m$ perhaps results from two distinct mechanisms -- first
is the appearance of a weak string like ordering around \pe{10}, and second is the formation
of hydroclusters\citep{wagner06} at high shear rates .
It is interesting to note that both of these mechanisms have the same apparent effect of causing 
a reduction in the flow alignment of the particles.

\section{Conclusions}\label{sec:conc} 
In this study we investigated the orientation and microstructure
in sheared Brownian suspensions of hard dicolloidal particles. 
Studies were conducted for a wide range of volume fractions ($0.3 \leq \phi \leq 0.5$)
and shear rates ($0 \leq Pe \leq 1000$) in suspensions of homonuclear particles (aspect ratio 1.1)
and fused-dumbbell particles (aspect ratio 1.5).
The microstructure was found to lack a long range order at all volume fractions,
though evidence of a weak string like ordering was apparent in suspensions
of homonuclear particles at high volume fractions and at intermediate to high shear rates.
The orientation distribution was found to be random at low shear rates.
At the higher shear rates, homonuclear particles exhibited enhanced
vorticity alignment at all volume fractions; fused-dumbbell
particle suspensions, in contrast, exhibited reduced vorticity alignment
at low volume fractions, and a negligible shift relative to the vorticity axis
at higher volume fractions. The orientation distribution in the
velocity--gradient plane indicated an increased flow
alignment with increasing volume fractions for both particle shapes;
flow alignment also increased with increasing shear rates at low volume fractions,
but not necessarily at higher volume fractions. These observations on the
orientation distribution were consistently resolved in the light of previous studies on
Brownian and non-Brownian suspensions.

\bibliography{references}

\begin{thebibliography}{38}
\providecommand{\natexlab}[1]{#1}
\providecommand{\url}[1]{\texttt{#1}}
\expandafter\ifx\csname urlstyle\endcsname\relax
  \providecommand{\doi}[1]{doi: #1}\else
  \providecommand{\doi}{doi: \begingroup \urlstyle{rm}\Url}\fi

\bibitem[van Blaaderen(2006)]{blaaderen06}
A.~van Blaaderen.
\newblock Colloids get complex.
\newblock \emph{Nature}, 439:\penalty0 545--546, 2006.

\bibitem[Yang et~al.(2008)Yang, Kim, Lima, and Yi]{yang08}
S.~Yang, S.~Kim, J.~Lima, and G.~Yi.
\newblock Synthesis and assembly of structured colloidal particles.
\newblock \emph{Journal of Materials Chemistry}, 18:\penalty0 2177--2190, 2008.

\bibitem[Glotzer and Solomon(2007)]{glotzer07}
S.C. Glotzer and M.J. Solomon.
\newblock Anisotropy of building blocks and their assembly into complex
  structures.
\newblock \emph{Nature Materials}, 6:\penalty0 557--562, 2007.

\bibitem[Mitragotri and lahann(2008)]{mitragotri08}
S.~Mitragotri and J.~lahann.
\newblock Physical approaches to biomaterial design.
\newblock \emph{Nature Materials}, 8:\penalty0 15--23, 2008.

\bibitem[Kumar and Graham(2012)]{kumar_rev}
A.~Kumar and M.~D. Graham.
\newblock Margination and segregation in confined flows of blood and other
  multicomponent suspensions.
\newblock \emph{Soft Matter}, 2012.
\newblock \doi{10.1039/C2SM25943E}.

\bibitem[Johnson et~al.(2005)Johnson, van Kats, and van Blaaderen]{blaaderen05}
P.M. Johnson, C.M. van Kats, and A.~van Blaaderen.
\newblock Synthesis of colloidal silica dumbbells.
\newblock \emph{Langmuir}, 21\penalty0 (24):\penalty0 11510--11517, 2005.

\bibitem[Mock et~al.(2006)Mock, Bruyn, Hawkett, Gilbert, and Zukoski]{mock06}
E.B. Mock, H.~De Bruyn, B.S. Hawkett, R.G. Gilbert, and C.F. Zukoski.
\newblock Synthesis of anisotropic nanoparticles by seeded emulsion
  polymerization.
\newblock \emph{Langmuir}, 22\penalty0 (9):\penalty0 4037--4043, 2006.

\bibitem[Kim et~al.(2006)Kim, Larsen, and Weitz]{weitz06}
J.W. Kim, R.J. Larsen, and D.A. Weitz.
\newblock Synthesis of nonspherical colloidal particles with anisotropic
  properties.
\newblock \emph{Journal of American Chemical Society}, 128\penalty0
  (44):\penalty0 14374--14377, 2006.

\bibitem[Kumar and Higdon(2011{\natexlab{a}})]{kumar_jor}
A.~Kumar and J.~J.~L. Higdon.
\newblock Dynamics of the orientation behavior and its connection with rheology
  in sheared non-brownian suspensions of anisotropic dicolloidal particles.
\newblock \emph{Journal of Rheology}, 55:\penalty0 581--626,
  2011{\natexlab{a}}.

\bibitem[Jeffery(1922)]{jeffery22}
G.B. Jeffery.
\newblock The motion of ellipsoidal particles immersed in a viscous fluid.
\newblock \emph{Proceedings of the Royal Society of London. Series A},
  102\penalty0 (715):\penalty0 161--179, 1922.

\bibitem[Bretherton(1962)]{bretherton62}
F.P. Bretherton.
\newblock The motion of rigid particles in a shear flow at low reynolds number.
\newblock \emph{Journal of Fluid Mechanics}, 14:\penalty0 284--304, 1962.

\bibitem[Leal and Hinch(1971)]{leal71}
L.~G. Leal and E.~J. Hinch.
\newblock The effect of weak brownian rotations on particles in shear flow.
\newblock \emph{Journal of Fluid Mechanics}, 46:\penalty0 685--703, 1971.

\bibitem[Hinch and Leal(1972)]{leal72}
EJ~Hinch and LG~Leal.
\newblock The effect of brownian motion on the rheological properties of a
  suspension of non-spherical particles.
\newblock \emph{Journal of Fluid Mechanics}, 52:\penalty0 683--712, 1972.

\bibitem[Brenner(1974)]{brenner74}
H.~Brenner.
\newblock Rheology of a dilute suspension of axisymmetric brownian particles.
\newblock \emph{International Journal of Multiphase Flow}, 1:\penalty0
  195--341, 1974.

\bibitem[Meng and Higdon(2008)]{higdon08b}
Q.~Meng and J.J.L. Higdon.
\newblock Large scale dynamic simulation of plate-like particle suspensions.
  part ii: Brownian simulation.
\newblock \emph{Journal of Rheology}, 52:\penalty0 37--65, 2008.

\bibitem[Egres et~al.(2006)Egres, Nettesheim, and Wagner]{wagner06}
R.G. Egres, F.~Nettesheim, and N.J. Wagner.
\newblock Rheo-sans investigation of acicular-precipitated calcium carbonate
  colloidal suspensions through the shear thickening transition.
\newblock \emph{Journal of Rheology}, 50:\penalty0 685--709, 2006.

\bibitem[Shaqfeh and Koch(1988)]{koch88}
E.~S.~G. Shaqfeh and D.~L. Koch.
\newblock The effect of hydrodynamic interactions on the orientation of
  axisymmetric particles flowing through a fixed bed of spheres or fibers.
\newblock \emph{Physics of Fluids}, 31\penalty0 (4):\penalty0 728--743, 1988.

\bibitem[Sierou and Brady(2002)]{brady2002}
A.~Sierou and J.F. Brady.
\newblock Rheology and microstructure in concentrated noncolloidal suspensions.
\newblock \emph{Journal of Rheology}, 46:\penalty0 1031--1056, 2002.

\bibitem[Leal(1975)]{leal75}
L.G. Leal.
\newblock The slow motion of slender rod-like particles in a second-order
  fluid.
\newblock \emph{Journal of Fluid Mechanics}, 69:\penalty0 305--337, 1975.

\bibitem[Kumar and Higdon(2010)]{kumar10_pre}
A.~Kumar and J.~J.~L. Higdon.
\newblock Origins of the anomalous stress behavior in charged colloidal
  suspensions under shear.
\newblock \emph{Physical Review E}, 82:\penalty0 051401, 2010.

\bibitem[Kumar(2010)]{kumar10_dis}
A.~Kumar.
\newblock \emph{Microscale dynamics in suspensions of non-spherical particles}.
\newblock PhD thesis, University of Illinois at Urbana-Champaign, 2010.

\bibitem[Sierou and Brady(2001)]{brady2001}
A.~Sierou and J.F. Brady.
\newblock Accelerated stokesian dynamics simulations.
\newblock \emph{Journal of Fluid Mechanics}, 448:\penalty0 115--146, 2001.

\bibitem[Kumar and Higdon(2011{\natexlab{b}})]{kumar_jfm}
A.~Kumar and J.~J.~L. Higdon.
\newblock Particle {M}esh {E}wald {S}tokesian dynamics simulations for
  suspensions of non-spherical particles.
\newblock \emph{Journal of Fluid Mechanics}, 675:\penalty0 297--335,
  2011{\natexlab{b}}.

\bibitem[Ball and Melrose(1997)]{melrose97}
R.C. Ball and J.R. Melrose.
\newblock A simulation technique for many spheres in quasi-static motion under
  frame-invariant pair drag and brownian forces.
\newblock \emph{Physica A}, 247:\penalty0 444--472, 1997.

\bibitem[Banchio and Brady(2003)]{banchio03}
A.~J. Banchio and J.~F. Brady.
\newblock Accelerated stokesian dynamics: Brownian motion.
\newblock \emph{The Journal of chemical physics}, 118:\penalty0 10323, 2003.

\bibitem[Schunk et~al.(2012)Schunk, Pierce, Lechman, Grillet, Weiss, Stoltz,
  Heine, et~al.]{schunk12}
P~R Schunk, F.~Pierce, J~B Lechman, A~M Grillet, H.~Weiss, C.~Stoltz, D~R
  Heine, et~al.
\newblock Performance of mesoscale modeling methods for predicting rheological
  properties of charged polystyrene/water suspensions.
\newblock \emph{Journal of Rheology}, 56:\penalty0 353--384, 2012.

\bibitem[Kern and Frenkel(2003)]{frenkel03}
N.~Kern and D.~Frenkel.
\newblock Fluid--fluid coexistence in colloidal systems with short-ranged
  strongly directional attraction.
\newblock \emph{The Journal of chemical physics}, 118:\penalty0 9882, 2003.

\bibitem[Russel et~al.(1989)Russel, D.A.Saville, and Schowalter]{russel89}
W.B. Russel, D.A.Saville, and W.R. Schowalter.
\newblock \emph{Colloidal Dispersions}.
\newblock Cambridge University Press, 1989.

\bibitem[Foss and Brady(2000)]{foss2000}
D.R. Foss and J.F. Brady.
\newblock Structure, diffusion and rheology of brownian suspensions by
  stokesian dynamics simulation.
\newblock \emph{Journal of Fluid Mechanics}, 447:\penalty0 167--200, 2000.

\bibitem[Saad(2003)]{saad}
Y.~Saad.
\newblock \emph{Iterative Methods for Sparse Linear Systems}.
\newblock Society for Industrial and Applied Mathematics, 2003.

\bibitem[Vega et~al.(1992)Vega, Paras, and Monson]{vega92b}
C.~Vega, E.P.A. Paras, and P.~A. Monson.
\newblock On the stability of the plastic crystal phase of hard dumbbell
  solids.
\newblock \emph{Journal of Chemical Physics}, 97\penalty0 (11):\penalty0
  8543--8548, 1992.

\bibitem[Speedy(1997)]{speedy97}
R.J. Speedy.
\newblock Pressure of the metastable hard-sphere fluid.
\newblock \emph{Journal of Physics Condensed Matter}, 9\penalty0 (41):\penalty0
  8591--8599, 1997.

\bibitem[Bolhuis and Frenkel(1997)]{bolhius97}
P.~Bolhuis and D.~Frenkel.
\newblock Tracing the phase boundaries of hard spherocylinders.
\newblock \emph{Journal of Chemical Physics}, 106\penalty0 (2):\penalty0
  666--687, 1997.

\bibitem[Brady and Morris(1997)]{brady97}
J.F. Brady and J.F. Morris.
\newblock Microstructure of strongly sheared suspensions and its impact on
  rheology and diffusion.
\newblock \emph{Journal of Fluid Mechanics}, 348:\penalty0 103--139, 1997.

\bibitem[Ackerson(1990)]{ackerson90}
B.~J. Ackerson.
\newblock Shear induced order and shear processing of model hard sphere
  suspensions.
\newblock \emph{Journal of rheology}, 34:\penalty0 553, 1990.

\bibitem[Chen et~al.(1994)Chen, Chow, Ackerson, and Zukoski]{chen94}
L.B. Chen, M.K. Chow, B.J. Ackerson, and C.F. Zukoski.
\newblock Rheological and microstructural transitions in colloidal crystals.
\newblock \emph{Langmuir}, 10:\penalty0 2817--2829, 1994.

\bibitem[Chow and Zukoski(1995)]{chow95}
M.K. Chow and C.F. Zukoski.
\newblock Nonequilibrium behavior of dense suspensions of uniform particles:
  Volume fraction and size dependence of rheology and microstructure.
\newblock \emph{Journal of Rheology}, 39\penalty0 (1):\penalty0 33--59, 1995.

\bibitem[Mock and Zukoski(2007)]{mock07}
E.B. Mock and C.F. Zukoski.
\newblock Investigating microstructure of concentrated suspensions of
  anisotropic particles under shear by small angle neutron scattering.
\newblock \emph{Journal of Rheology}, 51\penalty0 (3):\penalty0 541--559, 2007.

\end{thebibliography}
\end{document}